\documentclass[aps,prd,twocolumn,nofootinbib,showpacs,superscriptaddress,groupedaddress]{revtex4-1}

\usepackage{amsmath}
\usepackage{amssymb}
\usepackage{graphicx}
\usepackage{bm}
\usepackage{epsfig}
\usepackage{amsfonts}
\usepackage{color}
\usepackage{mathtools}
\usepackage{natbib}

\newcommand{\beq}{\begin{equation}}
\newcommand{\eeq}{\end{equation}}
\newcommand{\bea}{\begin{eqnarray}}
\newcommand{\eea}{\end{eqnarray}}
\newcommand{\beas}{\begin{eqnarray*}}
\newcommand{\eeas}{\end{eqnarray*}}

\newcommand{\Si}{\mathrm{Si}}
\newcommand{\mD}{m_{\mathrm{D}}}
\newcommand{\e}{0^+}

\newcommand{\bq}{{\bf q}}

\newcommand{\mO}{{\mathcal{O}}}

\newcommand{\Tint}[1]{{\hbox{$\sum$}\!\!\!\!\!\!\!\int\,}_{\!\!\!\!\raise-0.9ex\hbox{$\scriptstyle{#1}$}}}
\renewcommand{\Re}{\rm Re}
\renewcommand{\Im}{\rm Im}

\newcommand{\MeV}{\rm MeV}
\newcommand{\GeV}{\rm GeV}
\newcommand{\fm}{\rm fm}

\begin{document}

\title {A hard thermal loop benchmark for the extraction of the nonperturbative $Q\bar{Q}$ potential}

\author{Yannis Burnier}
\author{Alexander Rothkopf}
\affiliation{Albert Einstein Center for Fundamental Physics, Institute for Theoretical Physics, University of Bern, 3012 Bern, Switzerland}
\date{\today}

\begin{abstract}
The extraction of the finite temperature heavy quark potential from lattice QCD relies on a spectral analysis of the Wilson loop. General arguments tell us that the lowest lying spectral peak encodes, through its position and shape, the real and imaginary part of this complex potential. Here we benchmark this extraction strategy using leading order hard-thermal loop (HTL) calculations. I.e. we analytically calculate the Wilson loop and determine the corresponding spectrum. By fitting its lowest lying peak we obtain the real- and imaginary part and confirm that the knowledge of the lowest peak alone is sufficient for obtaining the potential. Access to the full spectrum allows an investigation of spectral features that do not contribute to the potential but can pose a challenge to numerical attempts of an analytic continuation from imaginary time data. Differences in these contributions between the Wilson loop and gauge fixed Wilson line correlators are discussed. To better understand the difficulties in a 
numerical extraction we deploy the Maximum Entropy method with extended search space to HTL correlators in Euclidean time and observe how well the known spectral function and values for the real and imaginary part are reproduced. Possible venues for improvement of the extraction strategy are discussed.
\end{abstract}
\maketitle

\section{Introduction}

 Twenty seven years ago Matsui and Satz \cite{Matsui:1986dk} proposed the melting of $J/\Psi$, i.e. the ground state of the $c\bar{c}$ vector channel, as signal for the deconfinement transition in heavy-ion collisions. The recent success of relativistic heavy-ion experiments \cite{Adare:2006ns,Tang:2011kr,Chatrchyan:2012np,Abelev:2012rv} in observing the relative suppression of charmonium and bottomonium serves as further motivation to develop a first principle description of the phenomena.

In the framework of effective field theories, heavy quarks can be described by non-relativistic quantum chromodynamics (NRQCD) obtained form QCD by integrating out the hard energy scale, given by the rest mass of the heavy quarks. To describe the bound state of two quarks, one can further integrate out the typical momentum exchange between the bound quarks (see \cite{Brambilla:2004jw} and references therein), which leads to potential non-relativistic QCD (pNRQCD).
In this effective field theory the bound state is described by a two point function satisfying a Schr\"odinger equation.

At zero temperature, the potential between a heavy quark and anti-quark is defined from the late time behavior of a Wilson loop and can be directly calculated in Euclidean-time lattice simulations or in perturbation theory. At small distances, where perturbation theory converges, both results agree \cite{Bazavov:2012ka}. 

At high temperature, above the QCD phase transition, one might first expect that the problem becomes simpler as the potential is not confining anymore. Actually, this is not the case since even a proper definition of the potential becomes non-trivial. In fact, the presence of a heat bath is most conveniently incorporated in a Euclidean time framework with finite temporal extend. There the Wilson loop depends on imaginary time and needs to be analytically continued to real time. Only from the large real-time, i.e. $t\to\infty$ behavior, the finite temperature potential can be extracted and happens to be complex \cite{Laine:2006ns,Brambilla:2008cx}. Its imaginary part can be interpreted as Landau damping \cite{Beraudo:2007ky} and describes the decaying correlation of the $Q\bar Q$ system with its initial state due to scatterings in the plasma. 

Along the lines presented in \cite{Laine:2006ns}, one can compute the potential in finite temperature perturbation theory. This is a demanding task, as resummations need to be carried out in order to cure infrared divergences. To this day the full result is known only to leading order, whereas a short distance expansion has been calculated to higher order \cite{Brambilla:2010vq,Brambilla:2011mk}. Even if higher orders were available, observing the deconfining transition will remain beyond the reach of perturbation theory.

In Ref.~\cite{Rothkopf:2011db}, a method was proposed to compute the heavy quark potential non-perturbatively from lattice QCD simulations. Starting from the measurement of the Euclidean Wilson loop on the lattice, its spectral function is reconstructed via the maximum entropy method (MEM). The definition of the potential is based on the peak structure of the Wilson loop spectrum. 

Previous numerical evaluations however lead to unexpected results: both the real and imaginary part appear to grow linearly at distances where other quantities, such as the free energies already show significant screening effects. This behavior persisted even at temperatures much larger than the QCD phase transition, where on general grounds, one would expect that the confining potential disappears because of Debye screening \cite{Digal:2005ht}. 

This problem was solved recently \cite{Burnier:2012az} by carefully disentangling the different timescales in the problem. Taking into account the remnants of early-time non-potential physics, the lowest lying spectral peak was found to deviate from a naive Lorentzian shape through skewing. Extracted values for real- and imaginary part based on this functional form result in a potential that is compatible with Debye screening. 

In this paper our aim is twofold: at first we wish to ascertain whether fitting of the lowest lying spectral peak indeed suffices to determine the static heavy quark potential, given the spectral function of the Wilson loop or even the gauge fixed Wilson line correlators. Subsequently it is our goal to better understand the challenges facing a numerical determination of the spectral function by Bayesian analytic continuation. Since in the perturbative approach both Euclidean correlator and spectrum are known, the outcome of the numerical reconstruction can be readily compared.


In section \ref{sect2} we review the basics of the method of Ref.~\cite{Rothkopf:2011db} and its improvement introduced in \cite{Burnier:2012az}, which form the basis of the extraction of the potential from lattice simulations. From calculations of the real-time Wilson loop as well as gauge fixed Wilson line correlators in section \ref{sect3} we determine and investigate the corresponding spectral functions in section \ref{sect4}. While in section \ref{sect5} we apply the peak fitting procedure of \cite{Burnier:2012az} to the HTL spectra, section \ref{sect6} scrutinizes how well these spectra can be obtained with the maximum entropy method from the HTL Euclidean correlators. Our conclusion in section \ref{sect7} discusses the limitations of the method and points toward further possible improvements.

\section{Heavy quark potential from Euclidean correlators}\label{sect2}

The description of the interactions between a pair of heavy quarks and antiquarks at finite temperature in terms of a quantum mechanical potential $V(r)$ requires the relevant physics to be well separated from the energy scale of pair creation. In particular 
\begin{align}
 \frac{\Lambda_{QCD}}{m_Q}\ll 1, \; \frac{T}{m_Q}\ll 1
\end{align}
needs to be fulfilled\footnote{See for instance \cite{Brambilla:2013dpa} for the discussion of the different limiting cases and their physics}, which is satisfied exactly in the static limit ($m_Q\to\infty$). In that case, the propagation amplitude of an infinitely heavy quark pair can be described by a rectangular temporal Wilson loop $W_\square(t,r)$ where $t,r$ are its temporal and spatial extend. This real-time quantity is defined as the closed contour integral over the matrix valued gauge field $A^\mu(x)=A^\mu_a(x) T^a$ along the path of the heavy quarks
\begin{align}
 W_\square(t,r)=\frac{1}{N_c}\,{\cal P} {\rm Tr}[{\rm exp}[-ig\oint_\square dx^\mu A_\mu(x)]].
\end{align}
If the scale hierarchy holds, it is permissible to substitute the field theoretical interactions by an instantaneous potential, so that $W_\square(t,r)$ obeys a Schr\"odinger type equation
\begin{align}
 i\partial_t W_\square(t,r) = \Phi(t,r) W_\square(t,r). \label{schro}
\end{align}
At late times, on expect the function $ \Phi(t,r)$ to become time independent, so that we may define the heavy quark static potential as
\begin{align}
 V(r)=\lim_{t\to\infty} \Phi(r,t). \label{Eq:DefPot}
\end{align}
 Due to the complex weighting factor in Feynman's path integral, we cannot calculate the real-time Wilson loop using lattice QCD Monte Carlo simulations. Instead we have to rely on an analytic continuation of Euclidean time quantities that are accessible by these numerical methods. In order to connect the heavy-quark potential $V(r)$ and the Euclidean Wilson Loop one introduces a spectral representation of the real-time quantity,
\begin{align}
 W_\square(r,t)=\int d\omega \; e^{-i\omega t}\; \rho_\square(r,\omega) \label{Eq:FourTrnasWL},
\end{align}
where the time dependence now resides entirely in the integral kernel. Note that the function $\rho_\square(r,\omega)$ is not just a Fourier transform but can be shown to be a positive definite spectral function \cite{Rothkopf:2009pk}\footnote{It is important to distinguish this $r$-dependent Wilson loop spectral function from the quarkonium spectral function \cite{Burnier:2007qm,Burnier:2008ia,Ding:2012iy,Ding:2012sp} representing the physical quarkonium spectrum.}. After analytic continuation $t=-i\tau$ one observes that only the integral kernel has changed, whereas the spectral function remains the same
\begin{align}
 W_\square(r,\tau)=\int d\omega \; e^{-\omega \tau}\; \rho_\square(r,\omega) \label{Eq:LapTransfWL}.
\end{align}
Using the Maximum Entropy Method (MEM), a form of Bayesian inference, it is in principle possible, albeit challenging, to invert Eq.\eqref{Eq:LapTransfWL} and thus to extract the spectral function from $W_\square(r,\tau)$. Note that the model independent method of refs.~\cite{Cuniberti:2001hm,Burnier:2011jq,Burnier:2012ts} is not directly applicable as the Wilson loop is not periodic. However a similar method could probably be developed from the general results of refs.~\cite{Viano1,Viano2}. Once we are in possession of the spectral function $\rho_\square$ we can insert Eq.\eqref{Eq:FourTrnasWL} into Eq.\eqref{Eq:DefPot}, which yields \cite{Rothkopf:2009pk}
\begin{align}
 V(r)=\lim_{t\to\infty} \frac{ \displaystyle \int d\omega \;\omega \; e^{-i\omega t} \; \rho(r,\omega)}{\displaystyle \int d\omega \; e^{-i\omega t} \; \rho(r,\omega)}\label{Eq:DefPotSpec}.
\end{align}
Direct application of this formula in the case of a numerically reconstructed spectral function is very difficult. It is however possible to determine those structures in the spectral function, which dominate the integral in the infinite time limit. 

If we suppose that the time independent potential description holds for all times $t$ i.e. $\Phi(t,r)=V(r)$ in equation (\ref{schro}), an intuitive connection between spectral features and the static potential can be established. In this case equation \eqref{schro} can be solved and the spectral function turns out to be a simple Breit Wigner peak
\beq
\rho(\omega,r)=\frac{{\rm Im}[ V](r)}{{\rm Im}[ V](r)^2+({\rm Re}[V](r)-\omega)^2}\label{BW},
\eeq
characterized by its peak position $\omega_0 (r)={\rm Re}[ V](r)$ and width $\Gamma_0(r)={\rm Im}[ V](r)$.

In general the function $\Phi(t,r)$ however is time dependent at early times and one expects that a wealth of structures, different from the simple Lorentzian example, exists in the spectrum of the Wilson loop at finite temperature. Note that if the potential description is ultimately applicable, the function $\Phi(r,t)$ will become time independent at late times and therefore a corresponding well defined lowest peak must exist. This part of the spectrum encodes all the relevant information on the potential and it alone needs to be reconstructed from the Euclidean correlator.

In Ref.~\cite{Rothkopf:2011db} it was assumed that the lowest peak is solely described by the late time behavior of the potential and is not affected by the time dependence of the potential at short times. It was shown in Ref.~\cite{Burnier:2012az} that this is actually not the case. The short time dynamics (non-potential terms, bound state formation) doesn't just create additional structures at high frequency but also significantly modifies the shape of the low frequency peak. The most general form of this low peak, derived in Ref.~\cite{Burnier:2012az}, can be written as
\begin{widetext}
\begin{eqnarray}
\notag \rho_\square(r,\omega)&=&\frac{1}{\pi}e^{{\rm Im}[\sigma_\infty](r)} \frac{|{\rm Im}[V](r)|{\rm cos}[{\rm Re}[\sigma_\infty](r)]-({\rm Re}[V](r)-\omega){\rm sin}[{\rm Re}[\sigma_\infty](r)]}{ {\rm Im}[V](r)^2+ ({\rm Re}[V](r)-\omega)^2}\\&&+c_0(r)+c_1(r)t_{Q\bar Q}({\rm Re}[V](r)-\omega)+c_2(r)t_{Q\bar Q}^2({\rm Re}[V](r)-\omega)^2+\cdots\label{Eq:FitShapeFull}
\end{eqnarray}
\end{widetext}
Note that this result can also be obtained from pNRQCD where $Re[\sigma_\infty]$ arises from the phase of the singlet normalization factors $Z_s^{(0)}(r)$ \cite{Brambilla:2004jw}

In order to calculate the potential $V(r)$ from Euclidean correlators we thus need to carry out the following steps:
\begin{enumerate}
 \item Calculate the Wilson loop $W_\square(r,\tau)$ at several separation distances $r$ for all possible values along the imaginary time axis $\tau\in[0,\beta]$.
 \item Use Bayesian inference to extract the most probable spectrum $\rho_\square(r,\omega)$ for each value of $r$.
 \item Use Eq.~\eqref{Eq:DefPotSpec} to determine the potential $V(r)$
 \begin{enumerate}
   \item by direct Fourier transform of the full $\rho_\square(r,\omega)$, which is usually impractical due to the uncertainties introduced by the MEM OR
   \item by fitting the lowest lying peak with the functional form~(\ref{Eq:FitShapeFull}) and analytically carrying out the Fourier transform in Eq.~\eqref{Eq:DefPotSpec}
 \end{enumerate}
\end{enumerate}

In the following section \ref{sect2} we prepare a testing ground for this extraction strategy based on analytic calculations of the real-time and Euclidean Wilson loop in the HTL resummed perturbative approach. Since the analytic continuation can be performed explicitly in HTL, item three of the above list can be tested independently from questions arising from possible inadequacies of the maximum entropy method. The availability of both the spectrum and Euclidean data points on the other hand furthermore allows us to check the degree of success of the MEM itself in the form of a realistic mock data analysis.

\section{Correlators from HTL resummed perturbation theory}\label{sect3}

\subsection{Wilson loop}\label{partWL}

In perturbation theory, the Wilson loop is calculated as an expansion in the coupling:
\beq
W_\square(\tau,r)=W_\square^{(0)}(\tau,r)+g^2 W_\square^{(2)}(\tau,r)+\mO(g^4),
\eeq
starting form $W_\square^{(0)}=1$. The first non-trivial term ($W_\square^{(2)}$) contains only a one gluon exchange and is not enough to describe the correct physics for large Euclidean time $\tau$. To improve this situation, we resort to the usual 'exponential' resummation \cite{Beraudo:2007ky}, noticing that
\beq
\log(W_\square(\tau,r))=g^2W_\square^{(2)}(\tau,r)+\mO(g^4).
\eeq
Thus a better approximation for $W_\square(\tau,r)$ is
\beq
W_\square(\tau,r)=\exp(g^2W_\square^{(2)}(\tau,r))+\mO(g^4),
\eeq
as it resums all 'ladder diagrams' and contains the correct leading order ($g^2$) large $\tau$ behavior.

\subsubsection{Leading order term}

We now turn to the calculation of $W_\square^{(2)}(\tau,r)$, for which we set the $r$ direction along the third spatial axis. In hard thermal loop (HTL) resummed perturbation theory, all diagrams contributing to $W_\square^{(2)}(\tau,r)$ have one HTL gluon running between the lines of the Wilson loop \cite{Laine:2006ns}:
\bea
W_{\square}^{(2)}&=&C_F T\int \frac{d^3q}{(2\pi)^3}\frac{e^{i q_3 r}+e^{-iq_3 r}-2}{2}\biggl\{\notag\\&&\tau^2\Delta_{00}(0,q)+\sum_{q_0\neq0}(2-e^{iq_0\tau}-e^{-iq_0\tau})\label{Wpropa}\\&&\notag\times\biggl(\frac{2\Delta_{03}(q_0,q)}{q_0 q_3}+\frac{\Delta_{33}(q_0,q)}{ q_3^2}+\frac{\Delta_{00}(q_0,q)}{q_0^2}\biggr)\biggr\}.
\eea
The gluon HTL propagator, written in Euclidean space ($Q^2=q_i^2+q_0^2$) and covariant gauge reads:
\bea
\Delta_{\mu\nu}(Q) &=& \notag
\delta^{ab} \Tint{Q} e^{i Q(x - y)}
\Biggl[\frac{P^T_{\mu\nu}(Q)}{Q^2 + \Pi_T(Q)}  \\&&+
\frac{P^L_{\mu\nu}(Q)}{Q^2 + \Pi_L(Q)} + \xi \frac{q_\mu q_\nu}{(Q^2)^2} 
\Biggr],
\eea
while the HTL self-energies $\Pi_{E,T}$ are given in Appendix \ref{A} and the projectors take the form:
\bea
P^T_{00}(Q) & = &\notag
P^T_{0i}(Q) = P^T_{i0}(Q)= 0, \quad P^T_{ij}(Q) = \delta_{ij} - \frac{ q_i q_j}{\bq^2},
\\P^L_{\mu\nu}(Q)& = &\delta_{\mu\nu} - \frac{q_\mu q_\nu}{Q^2}- P^T_{\mu\nu}(Q).
\eea
Following Ref.~\cite{Laine:2006ns}, we rewrite the HTL self-energies as spectral functions,
\beq
\frac{1}{q_0^2+\bq^2+\Pi_{L,T}(q_0,\bq)}=\int_{-\infty}^\infty \frac{dq^0}{\pi}\frac{\rho_{L,T}(q^0)}{q^0-iq_0},\label{spectralf}
\eeq
so that we can perform the sum over $q_0$ analytically:
\bea
W_\square^{(2)}(\tau,r)&=& C_F \int \frac{d^3\bq}{(2\pi)^3} \frac{e^{i q_3 r}+e^{-iq_3 r}-2}{2}\Biggl\{\\&&\frac{\tau}{\bq^2+\Pi_L(0,\bq)}+\int_{-\infty}^{\infty} \frac{dq^0}{\pi}n_B(q^0)h(\tau,q^0)\notag
\\ &&\times\notag
\Biggl[\rho_L(q^0,\bq)\left(\frac{1}{\bq^2}-\frac{1}{(q^0)^2}\right)\\&&+\rho_T(q^0,\bq)\left(\frac{1}{q_3^2}-\frac{1}{\bq^2}\right)\Biggr]\Biggr\},\notag
\eea
where we abbreviated the $\tau$ dependence of the second term through the function
\beq 
h(\tau,q^0)=1+e^{\beta q^0}-e^{\tau q^0}-e^{(\beta-\tau)q^0}.
\eeq
We can write the spatial vector $\bq$ in spherical coordinates $(q=|\bq|,\theta,\phi)$ and $q_3= q \cos\theta$.
In an isotropic plasma, the HTL spectral functions and self-energies depend on $q,q^0$ only.
Integrating over $\phi$ is trivial and the integral over $c =\cos\theta$ involves
\bea
\int_{-1}^1\frac{e^{i q r c}+e^{-iq r c}-2}{2} dc&=&2\left(\frac{\sin(q r)}{q r}-1\right),\\
\int_{-1}^1\frac{e^{i q r c}+e^{-iq r c}-2}{2 c^2} dc&=&2\left(1-\cos(q r)- r q\Si(q r) \right),\notag
\eea
where $\Si$ is the sin integral function. Performing the angular integrals and using $\Pi_E(0,\bq)=\mD^2$ gives
\bea
&&W_\square^{(2)}(\tau,r)=C_F\int_0^\infty \notag \frac{dq}{2\pi^2}~\tau\; \frac{q^2}{q^2+\mD^2}\left(\frac{\sin(qr)}{qr}-1\right)\\\notag
&&\quad+\int_{-\infty}^\infty \frac{dq^0}{\pi}\int_0^\infty \frac{dq}{2\pi^2} n_B(q^0)h(\tau,q^0)
\\&&\quad\times\Big\{\left(\frac{\sin(qr)}{qr}-1\right)\left(1-\frac{q^2}{(q^0)^2}\right)\rho_L(q^0,q)
\\&&\quad+\left(2-\frac{\sin(qr)}{qr}-\cos(qr)-qr\Si(qr)\right)\rho_T(q^0,q)\Big\}.\notag\label{cein}
\eea
The first line of equation (\ref{cein}) is linear in $\tau$, whereas the next lines are proportional to $h(\tau,q^0)$ and therefore symmetric around $\tau=\beta/2$. We will consider these terms separately in the following:
\beq
W_\square^{(2)}(\tau,r)=W^{(2)}_{lin}(\tau,r)+W^{(2)}_{sym}(\tau,r).\label{Wsquare}
\eeq

\subsubsection{Part linear in $\tau$}\label{LinPWL}

The part linear in $\tau$ is formally divergent. Using dimensional regularization, the result can be read off from Ref.~\cite{Laine:2006ns}; the first line of equation (\ref{cein}) hence gives:
\bea
W^{(2)}_{lin}(\tau,r)&=&C_F\int_0^\infty \frac{dq}{2\pi^2}~\tau\;\notag \frac{q^2}{q^2+\mD^2}\left(\frac{\sin(qr)}{qr}-1\right)\\&=&\frac{\tau C_F}{4\pi}\left(\frac{e^{-\mD r}}{r}+\mD\right). \label{clin}
\eea
In the limit $\tau\to it\to i\infty$ this part yields the real part of the potential:
\bea
{\rm Re}[ V](r)&=&g^2 \lim_{t\to\infty} i\frac{\partial}{\partial t} W^{(2)}_{lin}(it,r)\\&=&-\frac{g^2 C_F}{4\pi}\left(\frac{e^{-\mD r}}{r}+\mD\right).\notag
\eea
Note that the result is finite (for $r\neq0$) and the divergence at $r=0$ reflects the behavior of the Coulomb potential. 

On the lattice, this term behaves differently\footnote{The difference with dimensional regularization can be traced back to an infinite constant that is removed in the dimensional regularization procedure.}. Roughly speaking, the integral is truncated by the lattice cutoff $q<\Lambda$ and thus finite. In this case it is easy to see that it vanishes at $r=0$, which is expected as a Wilson loop without area is equal to unity. For $r>0$, it decreases quickly and formally goes to $-\infty$ in the limit of an infinite cutoff.

This behavior cannot be canceled by the other terms in equation (\ref{cein}) as they have a different $\tau$ dependence. It should also not be removed as it encodes the Coulomb part of the potential that we want to obtain. 
To make a connection to the lattice, we therefore introduce a UV cut-off, mimicking the finite lattice spacing. In this case, performing the integral over the momentum $q$ from zero to $\Lambda$ in equation (\ref{clin}) gives:
\beas
&&W^{(2),\Lambda}_{lin}(\tau,r)=C_F\frac{\tau}{2\pi^2}\biggl[-\Lambda +\mD \tan ^{-1}\left(\frac{\Lambda }{\mD}\right)\\&&+\frac{\cosh (\mD r) (\text{Si}(r(i \mD - \Lambda) )-\text{Si}(r\Lambda+ir \mD ))}{2 r}\\&&-\frac{(\pi -i \text{Ci}(r\Lambda -ir \mD))+i \text{Ci}(ir \mD+\Lambda r ))
   \sinh (\mD r)}{2 r}\biggr],
\eeas
where $\text{Si},\text{Ci}$ are the $\sin$ and $\cos$ integral function.
From the UV regularized version of the correlator we get the following potential,
\beq
{\rm Re}[ V^\Lambda](r)=g^2 \lim_{t\to\infty} i\frac{\partial}{\partial t} W^{(2),\Lambda}_{lin}(it,r),
\eeq
which is plotted in Fig.~\ref{Fig:RealImagPotHTL} together with the continuum ($\Lambda\to\infty$) potential.

\subsubsection{Symmetric part}\label{SymPartW}

We calculate here the symmetric part of the correlator $W^{(2)}_{sym}(\tau,r)$ containing the lines 2-4 of equation (\ref{cein}). 
The functions $\rho_{L,T}(q^0,q)$ receive a contribution from the cuts of $\Pi_{L,T}$ if $q>|q^0|$. For the opposite case $|q^0|>q$ they vanish except for a $\delta$-function contribution coming from the pole of $\Pi_{L,T}$.
In the following we calculate the contribution from the cuts and poles of the transverse and longitudinal self-energy separately,
\beq
W^{(2)}_{sym}=W^{(2)}_{cut}+W^{(2)}_{pole,L}+W^{(2)}_{pole, T}.\label{Wsym}
\eeq
As before we introduce a cutoff on the momentum to mimic the effects of the lattice regularization.

\paragraph{Cut contributions}
Using the symmetry $q_0\leftrightarrow -q_0$, the cuts contribute to the Euclidean Wilson loop as
\bea
&&W^{(2)}_{cut}(\tau,r)=C_F\int_0^\Lambda \frac{dq}{\pi^2} \int_{0}^q\notag \frac{dq^0}{\pi} n_B(q^0)h(\tau,q^0)\\&&\quad\times\Big\{\left(\frac{\sin(qr)}{qr}-1\right)\left(1-\frac{q^2}{(q^0)^2}\right)\rho_L(q^0,q)\label{Wcut}
\\&&\quad+\left(2-\frac{\sin(qr)}{qr}-\cos(qr)-qr\Si(qr)\right)\rho_T(q^0,q)\Big\},\notag 
\eea
where the integrals should be performed numerically and the functions $\rho_{L,T}$ are given in Appendix \ref{A}.  Note that in Eq.~\eqref{Wcut}, the limit $\Lambda\to\infty$ is well defined. 

\paragraph{Pole contribution form the longitudinal spectral function}
We can write the part of (\ref{cein}) coming from the pole contribution of the electric spectral function as:
\bea
W^{(2)}_{pole,L}(\tau,r)\notag&=&C_F\int_0^\Lambda \frac{dq}{\pi^2}\int_q^\infty dq^0\notag n_B(q^0)h(\tau,q^0)\\&&\notag\times\left(\frac{\sin(qr)}{qr}-1\right)\left(1-\frac{q^2}{(q^0)^2}\right)\delta(f_L(q^0))\\&\notag=&C_F\int_0^\Lambda \frac{dq}{\pi^2}n_B(q^0_L)h(\tau, q^0_L)\frac{1}{|f_L'(q^0_L)|}\\&&\times\left(\frac{\sin(qr)}{qr}-1\right)\left(1-\frac{q^2}{(q^0_L)^2}\right).\label{WpoleL}
\eea
Here $q^0_{L,T}$ is the solution of $f_{L,T}(q^0)=0,~q^0>0$ and the remaining integral is performed numerically.
The limit $\Lambda\to\infty$ also exists in this case (see Appendix \ref{pole}).

\paragraph{Pole contribution from the transverse spectral function}\label{TransWL}
We proceed in a similar way for the transverse spectral function.
\bea
W^{(2)}_{pole,T}(\tau,r)&=&C_F\int_0^\Lambda \frac{dq}{\pi^2}n_B( q^0_T)\frac{h(\tau, q^0_T)}{|f_T'(q^0_T)|}\label{31}\\&&\times\left(2-\frac{\sin(qr)}{qr}-\cos(q r)-q r\Si(qr) \right).\notag
\eea
Here, the limit $\Lambda\to\infty$ does not exist the integral in equation (\ref{31}) is linearly divergent (see Appendix \ref{pole}). 
Note that such divergences were already observed in \cite{Burnier:2009bk, Berwein:2012mw}, where the Wilson loop of maximal time extend $\tau=\beta$ is shown to diverge at next to leading order. The leading order divergence found in Eq. \ref{31} has yet a different nature and consistently vanishes for $\tau=0,\beta$. In dimensional regularization, it can be shown (see appendix \ref{C}) to match the cusp divergence \cite{Korchemsky:1987wg, Brandt:1982gz}, which in this case gives $\frac{ C_F g^2}{2 \pi^2\epsilon}$ \cite{Berwein:2012mw}.

Here, we are not interested in trying to renormalize the Wilson loop. It is not needed for our purposes as we aim at a comparison with lattice results, which are also not renormalized.
It is however interesting to note that these cusp divergences do not contribute to the potential and only make the Wilson loop heavily suppressed for $\tau\neq 0,\beta$, hence harder to measure with high accuracy. Removing these divergences in the lattice measurements, without affecting the potential would be of great help to improve the accuracy of the lattice data. One strategy deployed to this end could be the smearing of gluonic links \cite{Bazavov:2013zha}.

\subsubsection{Imaginary part of the potential}

From the symmetric part, we obtain the imaginary part of the potential,
\beq
i {\rm Im}[ V]^\Lambda(r)=g^2 \lim_{t\to\infty} i\frac{\partial}{\partial t} W^{(2),\Lambda}_{sym}(it,r).
\eeq
As in the end the infinite time limit will be taken, it is sufficient to consider the low frequency part of the $q^0$ integrals,
\bea
{\rm Im}[ V]^\Lambda(r)&=&g^2\notag \lim_{t\to\infty}\frac{\partial}{\partial t}\int_0^\Lambda \frac{dq}{\pi^2} \int_{0}^q \frac{dq^0}{\pi}  n_B(q^0) \frac{q^2}{(q^0)^2}\\&&\times h(it,q^0)
\left(\frac{\sin(qr)}{qr}-1\right)\rho_L(q^0,q)\notag
\eea
Performing the time derivative, using equation (\ref{rhoL}) and approximating $n_B(q^0)\approx T/q^0$ for small $q^0$ as well as the identity 
\beq
\lim_{t\to\infty}\frac{e^{it q^0} -e^{(\beta-it)q^0}}{q^0}=2\pi i \delta(q^0),
\eeq
we get:
\beq
{\rm Im}[ V]^\Lambda(r)=-\frac{g^2 C_F}{4\pi}\int_0^\Lambda\left(1-\frac{\sin(qr)}{qr}\right)\frac{2 q\mD^2}{(\mD^2+q^2)^2}dq\notag,
\eeq
which coincides with the expression obtained in \cite{Laine:2006ns,Beraudo:2007ky,Brambilla:2008cx}.

\subsubsection{Numerical evaluation}

To make close connection to actual lattice data with spatial lattice spacing $a=0.04{\rm \fm}$, we choose to fix the cut-off in our HTL calculations to 
\beq
\Lambda=\frac{\pi}{a},
\eeq
which naively corresponds to the largest momentum accessible under this finite resolution. Based on a numerical evaluation of the remaining integrals in eq. (\ref{Wsquare},\ref{Wsym}-\ref{31}), we can generate an arbitrary large number of datapoints spanning the imaginary time axis, which carry numerical errors of the order of the machine precision only. 

\begin{figure}
\includegraphics[width=6cm,angle=-90]{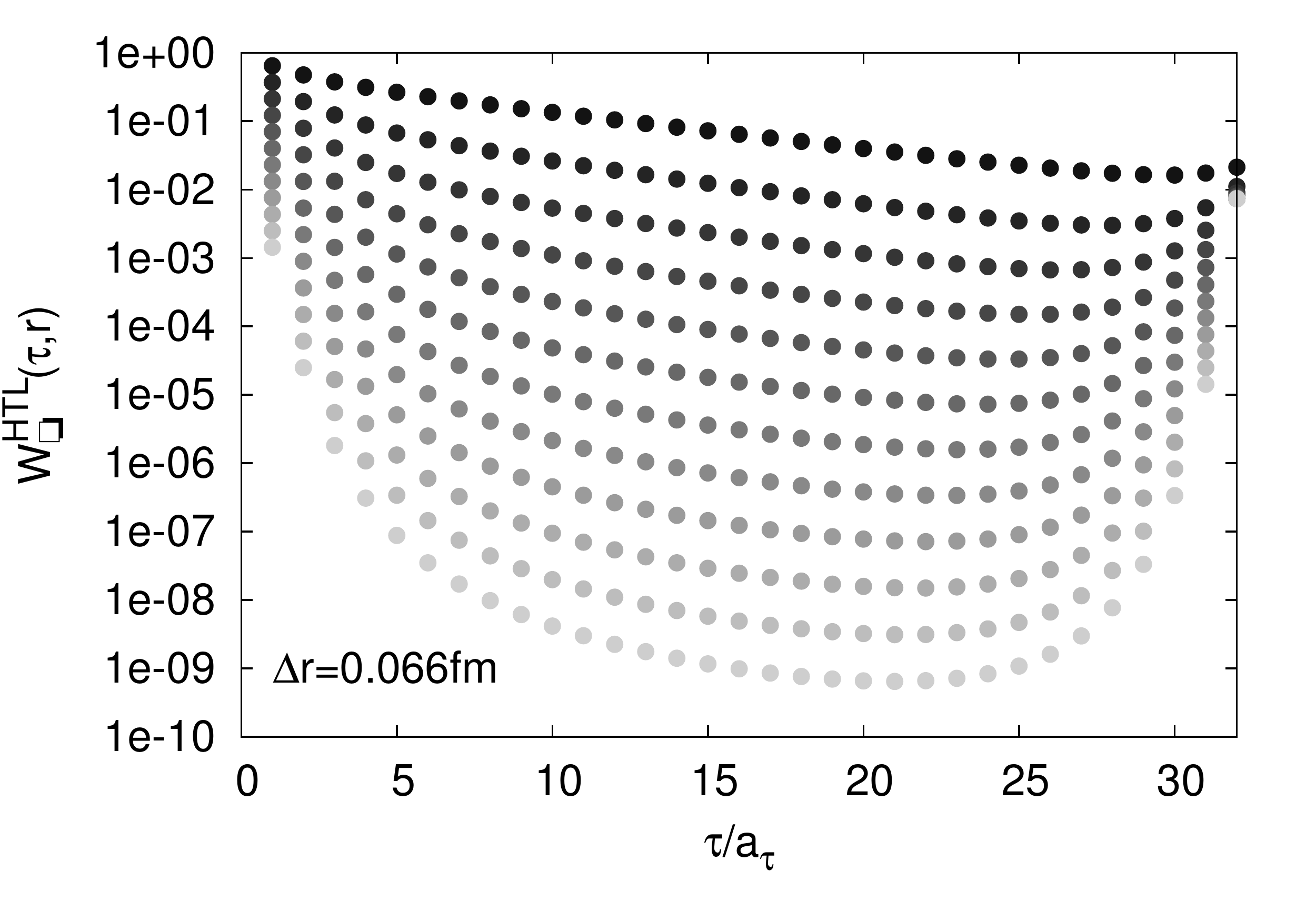}
\includegraphics[width=6cm,angle=-90]{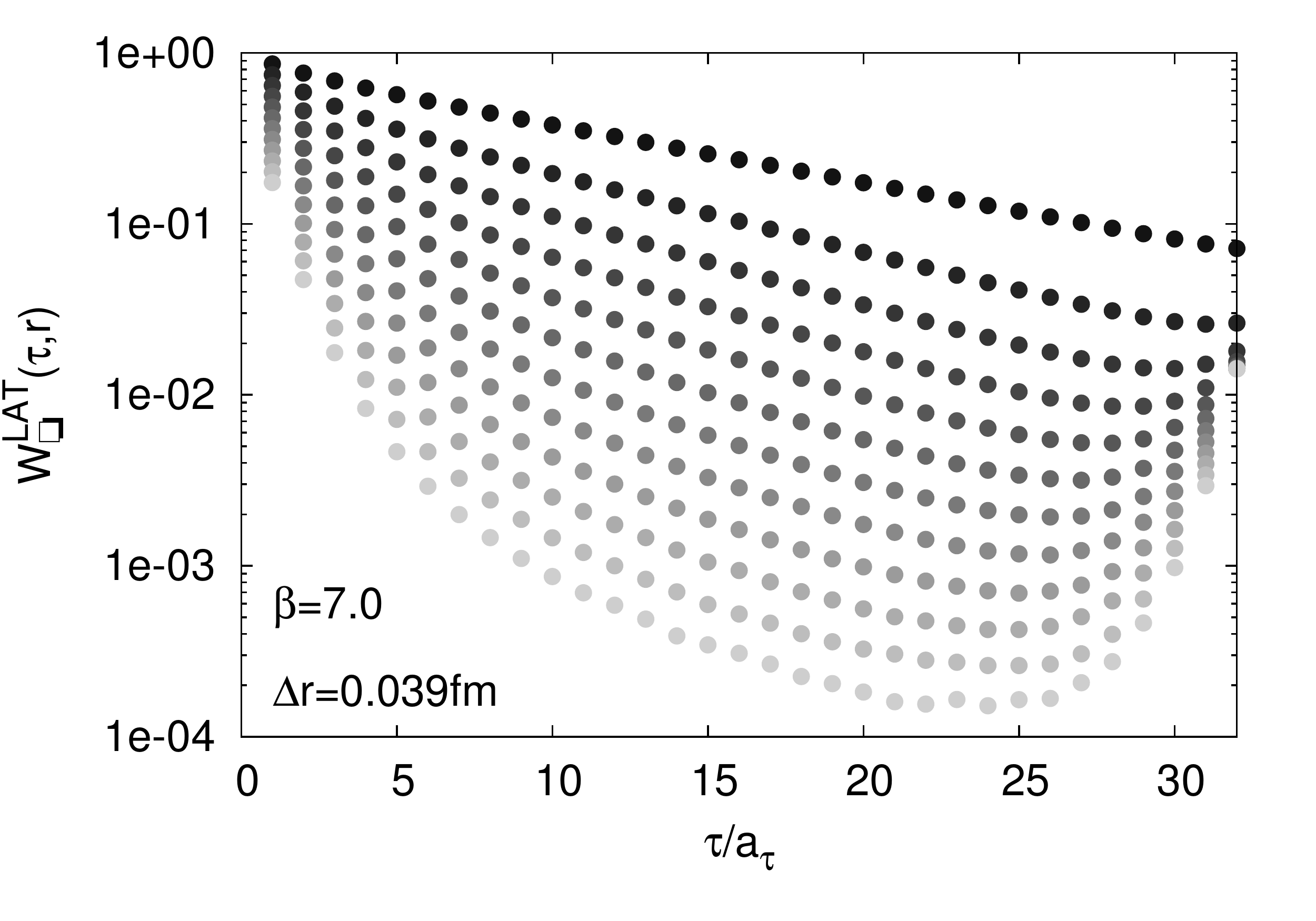}
\caption{(top) The Euclidean HTL Wilson loop $W^{\rm HTL}_\square(\tau,r)$ with momentum regularization $\Lambda=5\pi$ \GeV evaluated at $T=2.33\times 270$ \MeV in steps of $\Delta r=0.066\fm$. (bottom) Quenched lattice QCD Wilson loop from a lattice with $a=0.039\fm$ and anisotropy $\xi=4$ at $T=2.33T_C$.}
\label{1}
\end{figure}

Comparing this ideal HTL Euclidean regularized data to actual measurements from a Monte Carlo simulation in Fig.~\ref{1}, we find a strong qualitative resemblance. Both graphs exhibit three characteristic features, i.e. a suppression region at small $\tau$ together with an upward trend at $\tau\simeq \beta$, both of which are closely linked to the divergences observed in \ref{SymPartW}. The datapoints at intermediate $\tau$ are the ones encoding the potential. They exhibit nearly exponential behavior for small separation $r$, where also ${\rm Im}[V]$ is small but begin to show noticeable curvature for larger separation distances.

\begin{figure}
\includegraphics[width=9cm]{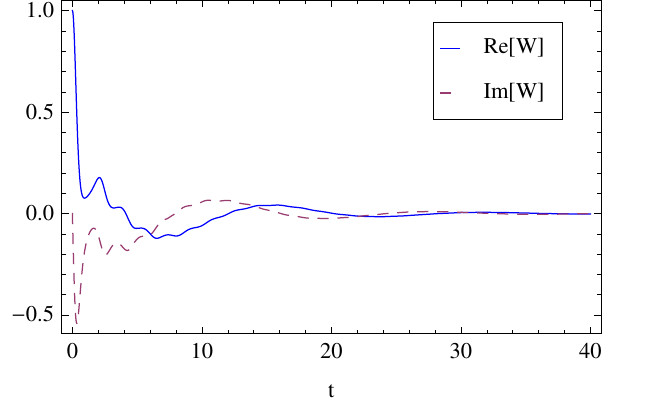}
\includegraphics[width=9cm]{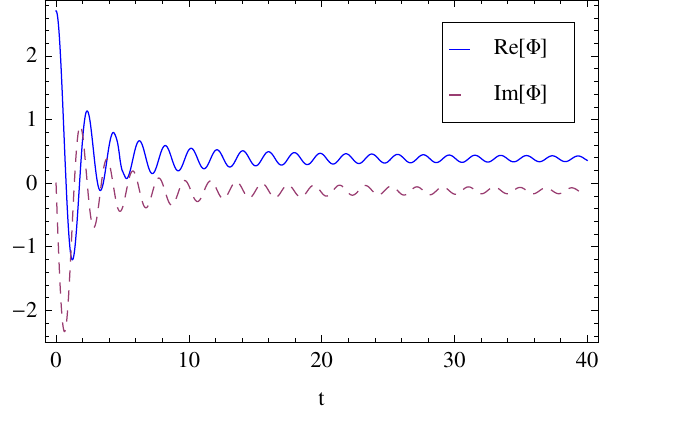}
\caption{(top) The HTL real-time Wilson loop $W^{\rm HTL}_\square(t,r)$ with momentum regularization $\Lambda=\pi$ \GeV evaluated at $r=1\GeV^{-1}$ and $T=2.33\times 270$ \MeV. (bottom) Time evolution of the quantity $\Phi(t,r)$ obtained from $W^{\rm HTL}_\square(t,r)$ through Eq.\ref{schro}.}
\label{2}
\end{figure}

After calculating the real-time values $W_\square(it,r)$ (see Fig.~\ref{2}) using a similar numerical evaluation of the integrals in (\ref{Wsquare},\ref{Wsym}-\ref{31}), it is possible to obtain the function $\Phi(t,r)$.  As shown in Fig.~\ref{2}, we can explicitly observe the approach of $\Phi(t,r)$ to a constant value and thus the emergence of a simple exponential behavior of the Wilson loop. Note that in Fig.~\ref{2} we show times $t<40\GeV^{-1}$ where the oscillatory behavior is clearly visible while a constant value is actually reached for larger $t$. We refrain from attaching any physical meaning to the length of the swing-in period, as it is dominated by the same cusp divergences that lead to the suppression of the Euclidean Wilson loop data points.

\subsection{Gauge fixed Wilson line correlator}

Cyclic Wilson line correlators (i.e.~color singlet Polyakov loops) fixed to Coulomb gauge have been extensively studied on the lattice, both for the determination of the zero temperature potential as well as in investigations into the free-energy difference between a medium with and without inserted heavy quarks (see for instance \cite{Maezawa:2011aa,Kaczmarek:2005ui}). Due to the absence of spatial Wilson lines connecting the temporal links, these quantities offer a significantly better signal to noise ratio than the Wilson loop, especially if the multilevel algorithm \cite{Luscher:1996ug} is applied. 

Besides the technical question of whether the removal of spatial connectors (or e.g. the application of smearing on spatial links) can lead to an improved lattice observable for the extraction of the potential, it is conceptually of interest to understand whether gauge independent information, such as the potential can be extracted from a gauge dependent quantity such as the Wilson line correlators\footnote{The crucial difference to potential models is that we do not investigate the single point $\tau=\beta$, but it is the full Euclidean time dependence of the gauge fixed correlator that is used to reveal the values of the potential.}. 

We proceed with the determination of the Euclidean time Wilson line correlator analogously to \ref{partWL}
\beq
W_{||}(\tau,r)=1+g^2 W_{||}^{(2)}(\tau,r)+\mathcal{O}(g^4).
\eeq
Calculating in leading order hard thermal loop (HTL) resummed perturbation theory we obtain the expression
\bea
W_{||}^{(2)}&=&C_F T\int \frac{d^3q}{(2\pi)^3}\frac{e^{i q_3 r}+e^{-iq_3 r}-2}{2}\biggl\{\tau^2\Delta_{00}(0,q)\notag\\&&+\sum_{q_0\neq0}(2-e^{iq_0\tau}-e^{-iq_0\tau})\frac{\Delta_{00}(q_0,q)}{q_0^2}\biggr\},\label{WLpropa}
\eea
which contains fewer terms than the Wilson loop of (\ref{Wpropa}).

\subsubsection{Coulomb gauge}

In Coulomb gauge, the HTL Euclidean gluon propagator reads
\bea
\Delta_{\mu\nu}(q_0,q) &=& \notag
\delta^{ab} \Tint{Q} e^{i Q(x - y)}
\Biggl[\frac{P^T_{\mu\nu}(Q)}{Q^2 + \Pi_T(Q)}  \\&&+
\frac{Q^2}{q^2}\frac{g_{\mu 0}g_{\nu 0}}{Q^2 + \Pi_L(Q)} 
\Biggr],
\eea
where the self energies $\Pi_{L,T}$ are the same as in covariant gauge (see Appendix \ref{A}).
Inserting the propagator into the expression (\ref{WLpropa}) for the Wilson line correlator gives
\bea
W_{||}^{(2)}&=& C_F T\int \frac{d^3{ \bf q}}{(2\pi)^3}\frac{e^{i q_3 r}+e^{-iq_3 r}-2}{2}\biggl\{\frac{\tau^2}{{ \bf q}^2+\mD^2}\notag\\&&+\sum_{q_0\neq0}\left[\frac{Q^2}{{ \bf q}^2q_0^2}\frac{2-e^{iq_0\tau}-e^{-iq_0\tau}}{Q^2+\Pi_L(Q)}
\right]\biggr\}.
\eea
We now rewrite the HTL self-energies as spectral functions, use the formulas collected in Appendix \ref{A} to perform the sum over $q_0$ and carry out the angular integrations:
\bea
W^{(2)}_{||}&=&C_F \int q^2\frac{dq}{2\pi^2} \left[\frac{\sin(qr)}{qr}-1\right] \Biggl\{\frac{\tau}{q^2+\mD^2}\\&&+\int_{-\infty}^{\infty} \frac{dq^0}{\pi}\left[\frac{1}{q^2}-\frac{1}{(q^0)^2}\right]\rho_L(q^0)n_B(q^0)h(\tau,q^0)\Biggr\}.\notag
\eea
We find that the Coulomb gauge Wilson line correlator features a similar structure as the Wilson loop
\bea
W^{(2)}_{||}(\tau,r) =W^{(2)}_{lin}(\tau,r)+\tilde{W}^{(2)}_{sym}(\tau,r)
\eea
the symmetric expression however being of much simpler form, depending only on the longitudinal HTL spectral function. At this point we can already anticipate that it is these terms present in both Wilson loop and Wilson line correlator, which contribute to the values of the potential. In particular the cusp singularity connected to the transverse spectral function identified in \ref{TransWL} is absent from the above expression. 

\subsubsection{Potential from the Wilson line correlator}

As in the case of the Wilson loop, a closed expression for the potential can be obtained using
\bea
V_{||}^{HTL}(r)&=&g^2 \lim_{t\to\infty} i\frac{\partial}{\partial t}W_{||}^{(2)}(it,r)\\\notag
&=&g^2\frac{C_F}{2\pi^2} \int dq \left[1-\frac{\sin(qr)}{qr}\right] \Biggl\{\frac{q^2}{\bq^2+\mD^2}\\&&\notag+\int_{-\infty}^{\infty} \frac{dq^0}{\pi}(q^2-(q^0)^2)\rho_L(q^0)n_B(q^0)\\&&\notag\hphantom{+\int_{-\infty}^{\infty} \frac{dq^0}{\pi}}\times\frac{e^{i t q_0}-e^{(\beta-i t) q_0}}{q^0}\Biggr\}.\notag
\eea
In the infinite time limit one can make use of 
\beq
\lim_{t\to\infty}\frac{e^{i t q_0}-e^{(\beta-i t) q_0}}{q^0}=2\pi i \delta(q^0),
\eeq
which leads us to the same result we encountered for the Wilson loop
\beq
 V^{HTL}_{||}(r)=-\frac{g^2 C_F}{4\pi} \Big[ m_D + \frac{e^{-m_Dr}}{r} - i T\phi(m_D r)\Big]\label{VHTL}
\eeq
with the imaginary part given by the integral expression
\begin{align}
 \phi(x)=2\int_0^\infty dz \frac{z}{(z^2+1)^2}\Big[1-\frac{\sin[zx]}{zx}\Big].
\end{align}

From a practical standpoint this result is encouraging, as it tells us that (to leading order in HTL) the information content regarding the potential encoded in the Coulomb gauge Wilson line correlator is the same as the one found in the Wilson loop. If such a relation persisted into the non-perturbative realm, the absence of cusp divergences and with it the improved signal to noise ratio would make this an ideal observable to reconstruct the potential.

\subsubsection{Numerical evaluation}

As for the Wilson loop we wish to compare the Euclidean HTL correlator to actual values measured in quenched lattice QCD Monte Carlo simulations. While the symmetric term $\tilde{W}^{(2)}_{sym}(\tau,r)$ is finite, the part linear in $\tau$ still requires a regularization. We deploy the same momentum space cutoff as introduced in \ref{LinPWL} and set its value to $\Lambda=5\pi\GeV$ in the following. 

\begin{figure}
 \begin{center}
 \includegraphics[width=6cm,angle=-90]{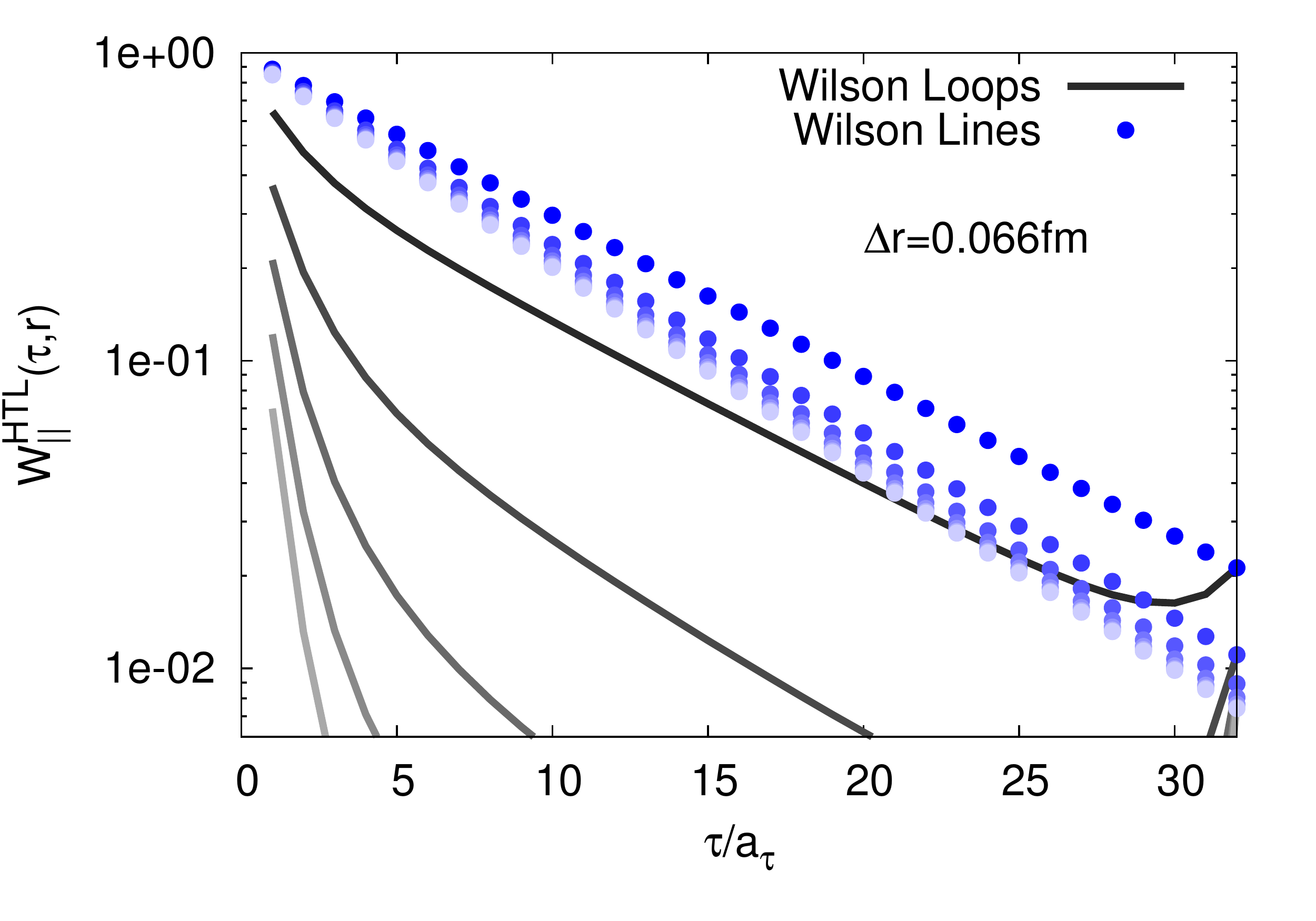}
  \includegraphics[width=6cm,angle=-90]{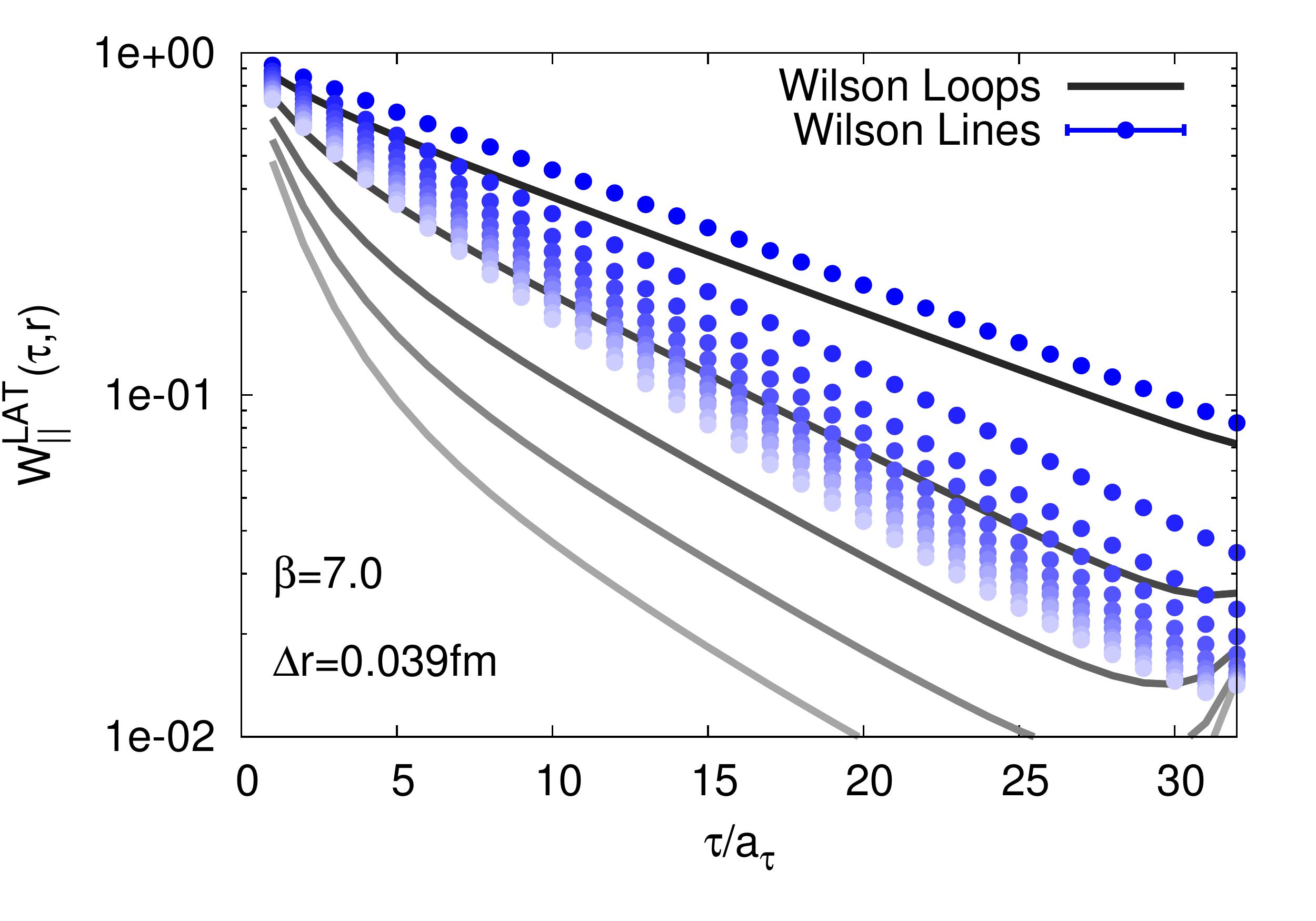}
  \caption{(top) The Euclidean time Coulomb gauge HTL Wilson line correlator $W^{\rm HTL}_{||}(\tau,r)$ with momentum regularization $\Lambda=5\pi$ \GeV evaluated at $T=2.33\times 270$ \MeV in steps of $\Delta r=0.066\fm$. (bottom) Quenched lattice QCD Wilson line correlator fixed to coulomb gauge from a lattice with $a=0.039\fm$ and anisotropy $\xi=4$ at $T=2.33T_C$. Note that contrary to the HTL result the two correlators do not agree at $\tau=\beta$ on the lattice.}\label{f2}
 \end{center}

\end{figure}

The absence of divergences in the symmetric part of the correlator leads to a significantly different behavior along the imaginary times $\tau$. As can be seen in the top graph in Fig.~\ref{f2}, where we plot the HTL Wilson line correlator and the first five HTL Wilson loops as comparison. The large suppression at early times as well as the upward trend near $\tau=\beta$ are almost absent. Hence most of the datapoints actually carry information on the potential. 

Interestingly in the case of the lattice QCD Wilson line correlator, the upward trend is still visible between the last and second to last time step. However contrary to the leading order HTL result, where $W_{||}^{\rm HTL}(\beta,r)=W_\square^{\rm HTL}(\beta,r)$ , the values of these two different correlators on the lattice do not agree at $\tau=\beta$.

\subsubsection{Covariant gauge}
The Wilson line correlator can be calculated in the covariant gauge as well. The result depends on the gauge parameter $\xi$, and contains additional end point divergences \cite{Aoyama:1981ev}. These terms however do not contribute in the infinite time limit so that the obtained potential is again the same as in the Wilson loop case (\ref{VHTL}). 

\section{Spectral functions from HTL resummed perturbation theory}\label{sect4}

\subsection{Spectrum of the Wilson loop}
The spectral function can be directly calculated from the real-time correlator via a Fourier transform,
\bea
\rho_\square(r,\omega)&=&\notag\frac{1}{2\pi}\int dt\; e^{i\omega t}W_\square(it,r)\\&=&\frac{1}{2\pi}\int dt\; e^{i\omega t}e^{g^2 W^{(2)}_\square(it,r)}+\mO(g^4).\label{fullrho}
\eea
We start by analytically investigating the low frequency behavior of this function, as it allows insight into the spectral structures that encode the physics of the heavy quark potential and will be used in its extraction in section \ref{sect5}. To benchmark the MEM extraction of the spectrum from Euclidean correlators it is however necessary to compare to the full spectrum, which we will determine from Eq.\eqref{fullrho} numerically.

\subsubsection{Analytical estimate for the low energy part of the spectral function}

Starting form equation (\ref{cein}), we introduce the momentum cutoff $\Lambda$ 
\beq
\rho^\Lambda_\square(r,\omega)=\frac{1}{2\pi}\int dt\;e^{i\omega t}e^{-i t{\rm Re}[ V]^\Lambda(r)}e^{\int_{-\infty}^\infty \frac{dq^0}{\pi}f_\square(q^0)},\label{rho}
\eeq
where the argument of the second exponential function reads
\bea
f_\square(q^0)&=&\notag g^2 C_F\int_0^\Lambda dq \frac{n_B(q^0)}{2\pi^2} h(it,q^0)\\&&\hspace{-5mm}\times\bigg\{\left(\frac{\sin(qr)}{qr}-1\right)\left(1-\frac{q^2}{(q^0)^2}\right)\rho_L(q^0,q)
\\&&\hspace{-5mm}+\left(2-\frac{\sin(qr)}{qr}-\cos(qr)-qr\Si(qr)\right)\rho_T(q^0,q)\bigg\}.\notag \label{f}
\eea
For small frequencies, the main contribution to the spectral function (\ref{rho}) comes form small values of $q^0$ in the above integral. Expanding equation (\ref{f}) around $q^0=0$ gives:
\bea
f_\square(q^0)&=&\frac{{\rm Im}[ V]^\Lambda(r)}{2\pi}\left[\frac{2-e^{itq^0}-e^{-itq^0}}{(q^0)^2}+\frac{e^{itq^0}-e^{-itq^0}}{2q^0}\right]\notag
\\&&+\mO\left((q^0)^0\right),
\eea
All terms with negative powers of $q^0$ are retained in this expression, as they dominate the integral for late times. Note that the imaginary part of the potential appears as an overall factor in the above expression. Within this approximation, the remaining integrals are carried out analytically and we get:
\bea
\rho^\Lambda_\square(r,\omega)&=&\frac{k}{2\pi}\Biggl(\frac{e^{i \frac{|{\rm Im}[ V](r)|}{2T}}}{|{\rm Im}[ V](r)|-i({\rm Re}[ V](r)-\omega)}\notag\\&&+\frac{e^{-i \frac{|{\rm Im}[ V](r)|}{2T}}}{|{\rm Im}[ V](r)|+i({\rm Re}[ V](r)-\omega)}\Biggr)
\\&=&\frac k\pi \frac{|{\rm Im}[ V](r)|\cos\delta_\square-({\rm Re}[ V](r)-\omega)\sin\delta_\square}{({\rm Im}[ V](r))^2+({\rm Re}[ V](r)-\omega)^2},\notag
\eea
with $\delta_\square=\frac{|{\rm Im}[ V](r)|}{2T}$ and $k$ denoting a not near specified normalization constant. 
From this result, we see that the pole of the spectral function indeed resides at $\omega={\rm Re}[ V](r)$ and the width of the peak is closely related to the imaginary part of the potential. The result however is not a Lorentzian, but is precisely of the form (\ref{Eq:FitShapeFull}) derived on general grounds in \cite{Burnier:2012az}. Note that the phase related to the skewing of the spectral peak is interestingly also given by the imaginary part of the potential 
\beq
{\rm Re}[ \sigma_\infty]=\delta_\square=|{\rm Im}[ V](r)|/2T.\label{ImVSq}
\eeq

\subsubsection{Full spectral function}

\begin{figure}
 \begin{center}
 \includegraphics[width=8.3cm]{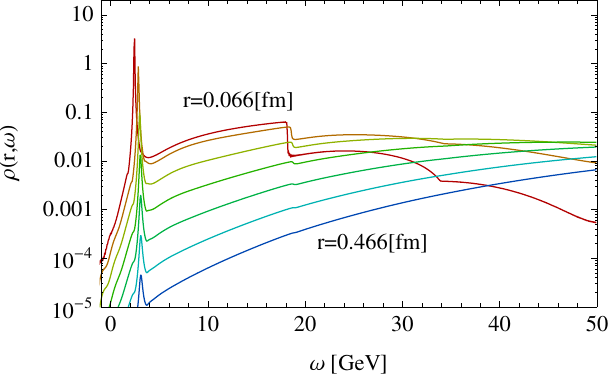}
  \caption{The HTL Wilson loop spectral function $\rho^\Lambda_\square(r,\omega)$ for different spatial separations $\Delta r=0.065$\fm . Note that the peak is extremely sharp but that its amplitude becomes very small at large $r$ in comparison to the huge background induced mostly by the cusp divergences.}
  \label{frhoW}
 \end{center}

\end{figure}

We proceed to calculate the full spectral function by integrating numerically equation (\ref{fullrho}). Applying the discrete Fourier transform to the real-time Wilson loop evaluated on a set of $N_t=25000$  points separated by a $\Delta t=\frac{1}{50}\frac{1}{{\rm \GeV}}$, we obtain its values for a wide range of frequencies partly shown in Fig.~\ref{frhoW}. 

As expected from the minute values of $\Im[V^{\rm HTL}]$ at small separation distances, the peak one finds is extremely sharp. However it also becomes clear that the amplitude of the peak is rapidly suppressed as $r$ increases. At the same time non-potential contributions related to the divergent terms in the symmetric part of $W^{(2)}(t,r)$ give rise to a huge background structure spanning a wide range of frequencies.

Note that at $\omega\approx18$ $\GeV$ a step in the otherwise smooth spectral function is visible. This is a manifestation of the momentum cutoff we introduced to regularize the formally divergent terms. At the same time one can observe that the spectrum continues beyond these frequencies, which is a reminder that the cutoff was not imposed on the HTL gluon spectral functions. 

In section \ref{sect5} we will use the fitting function (\ref{Eq:FitShapeFull}) to attempt an extraction of the heavy quark potential from the low frequency structures depicted in Fig.~\ref{frhoW}.

\subsection{Spectrum of the Wilson line correlator in Coulomb gauge}

Analogously we can obtain the the spectral function related to the real time Wilson loop correlator
\bea
\rho_{||}(r,\omega)&=&\notag\frac{1}{2\pi}\int dt\; e^{i\omega t}W_{||}(it,r)\\&=&\frac{1}{2\pi}\int dt\; e^{i\omega t}e^{g^2 W^{(2)}_{||}(it,r)}+\mO(g^4).\label{fullrhoL}
\eea
At leading order in the HTL resummed expansion, we again have:
\beq
\rho^\Lambda_{||}(r,\omega)=\frac{1}{2\pi}\int dt\;e^{i\omega t}e^{-i t{\rm Re}[ V]^\Lambda(r)}e^{\int_{-\infty}^\infty \frac{dq^0}{\pi}f_{||}(q^0)}
\eeq
with
\bea
f_{||}(q^0)&=& g^2 C_F\int_0^\infty  \frac{dq}{2\pi^2} n_B(q^0)h(it,q^0)\label{fL}\\&&\times\left(\frac{\sin(qr)}{qr}-1\right)\left(1-\frac{q^2}{(q^0)^2}\right)\rho_L(q^0,q)\notag.
\eea
The spectral function can then be calculated analytically close to its peak at small frequency, which yields
\bea
\rho^\Lambda_{||}(r,\omega)&=&
\frac1\pi \frac{|{\rm Im}[ V](r)|\cos\delta_{||}-({\rm Re}[ V](r)-\omega)\sin\delta_{||}}{({\rm Im}[ V](r))^2+({\rm Re}[ V](r)-\omega)^2}.\notag
\eea

Surprisingly at leading order in the HTL approximation we find that the skewing characterized by the quantity $\delta_{||}=\frac{|{\rm Im}[ V](r)|}{2T}$ is exactly the same as for the Wilson loop. Note that the same result can also be obtained in the covariant gauge.

\subsubsection{Full spectral function}

\begin{figure}
 \begin{center}
 \includegraphics[width=8.3cm]{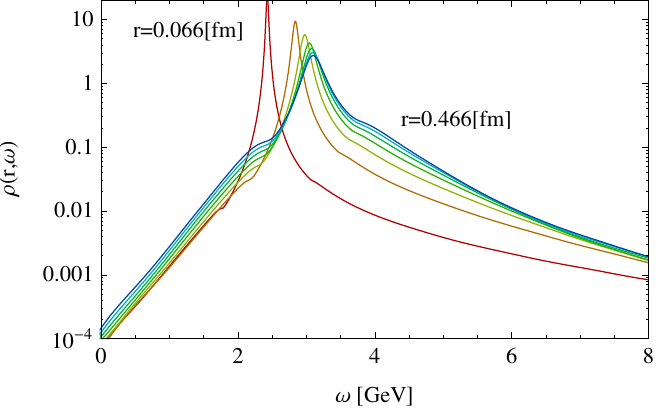}
  \caption{The spectral function of the HTL Wilson line correlator in Coulomb gauge $\rho^\Lambda_{||}(r,\omega)$ for different spatial separations $\Delta r=0.065$\fm . While the peak position, width and skewing is exactly as in the Wilson loop case (Fig.~\ref{frhoW}), the absence of the cusp divergences leads to a significantly reduced background and a much higher amplitude at larger separation distances. Note that the plotting range is much smaller than in Fig.~\ref{frhoW}.}
  \label{frhoC}
 \end{center}
 
\end{figure}

The full spectral functions for the HTL Wilson line correlator are plotted in Fig.~\ref{frhoC}. One immediately realizes from a comparison with Fig.~\ref{frhoW} that even though the peak position, width and skewing are equal to the Wilson loop case, the Coulomb gauge spectral function looks quite different. The first major difference is that the amplitude of the lowest lying peak depends much less on the separation distance $r$, the second is the virtual absence of the background terms populating a large frequency range in the Wilson loop case. Both facts are of course related, since their origin lies in the suppression of the Euclidean Wilson loop correlator induced in the presence of cusp divergences.

\section{The potential from perturbative spectral functions} \label{sect5}

Now that we are in possession of the full HTL spectra obtained from both the Wilson loop and the Wilson line correlator in Coulomb gauge, we can test whether the knowledge of the lowest lying spectral features alone suffices to reconstruct the values of the inter-quark potential in practice. To this end we fit the low $\omega$ region of $\rho^\Lambda(r,\omega)$ using the functional form (\ref{Eq:FitShapeFull}) and compare the extracted values with the analytically calculated $V^{\rm HTL}(r)$. We show here the fitting of the Wilson loop spectrum only, since its application to $\rho^\Lambda_{||}(r,\omega)$ gives exactly the same results (the potential and the skewing are the same). In section \ref{sect6}, where the numerical reconstruction of the spectra from Euclidean time correlator data is concerned, the differences in e.g. the background contributions will however play a major role.

\begin{figure}[t!]
\centering
\includegraphics[scale=0.95]{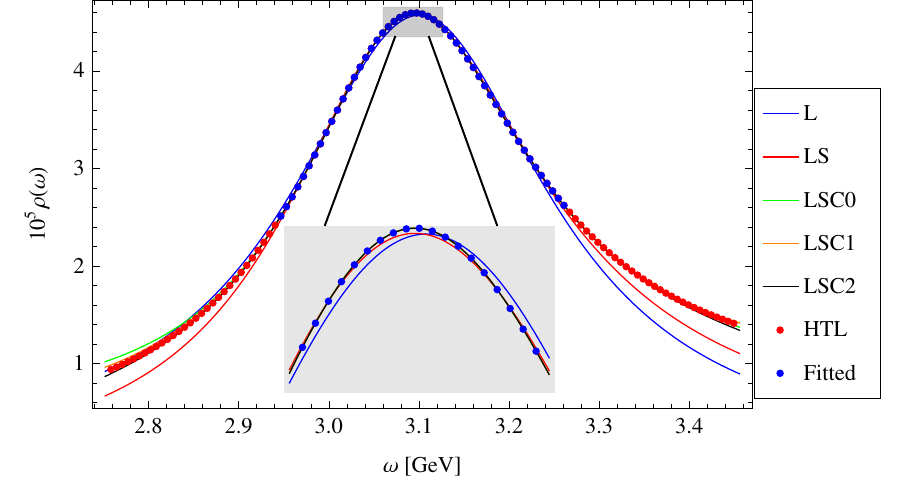}
 \caption{Fits to the UV regularized $(\Lambda=5\pi)$ HTL spectral function $\rho^\Lambda_\square(r,\omega)$ at  $r=0.49{\rm {\rm \fm}}$ (right) with a naive Lorentzian (L), a skewed Lorentzian (LS) and a skewed Lorentzian with additional polynomial  terms (LSC0,LSC1,LSC2). Note that only the blue points (labeled ''Fitted'') are used for the fit and hence only these points enter the determination of the potential.} \label{Fig:SpecFits}
\end{figure}

\begin{figure}[th!]
\centering
\includegraphics[scale=1.1]{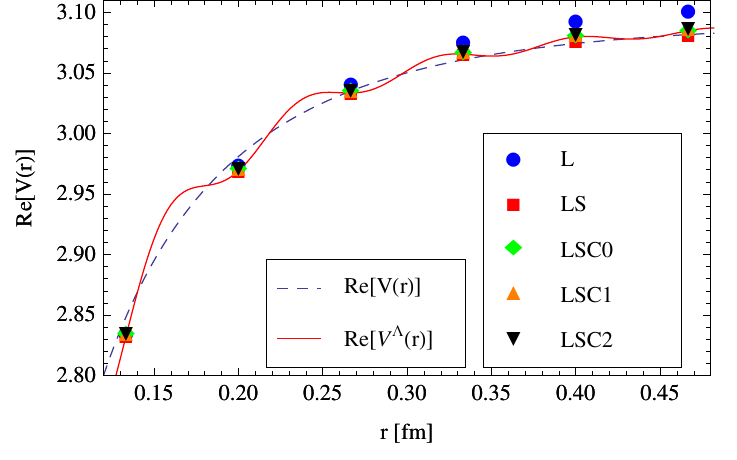}\hspace{0.5cm}
\includegraphics[scale=1.1]{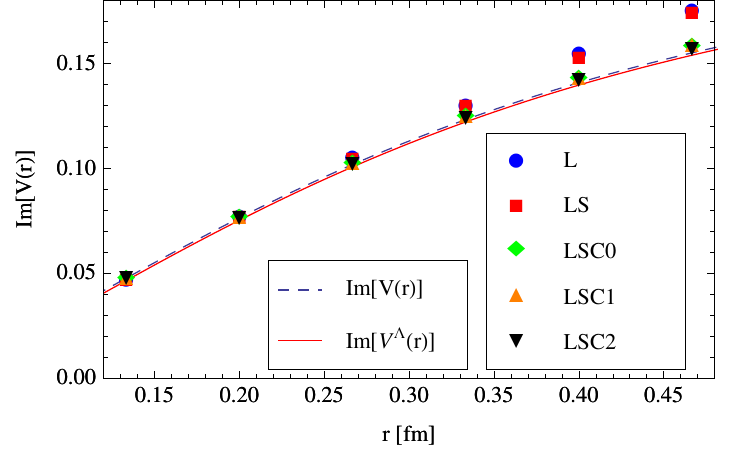}
 \caption{ (top) Real- and (bottom) imaginary  part of the UV regularized ($\Lambda=5\pi {\rm \GeV}$) HTL heavy quark potential (red, solid line) at $T=2.33T_C$ as well as the potential without cutoff (gray, dashed line). The various symbols denote the extracted values from fits of the  HTL Wilson loop spectra based on a Lorentzian (L), skewed Lorentzian (LS) and a skewed Lorentzian with background terms (LSC0,LSC1,LSC2). Note that the simple Lorentzian consistently overestimates the correct values. The determination of the real part suffers only slightly from a worsening of the fit ($LSC2\to LS$) but rough agreement is still visible.  On the other hand a successful extraction of the imaginary part requires at least the presence of the first background term (LSC0), once the width of the spectral peak lies above 150\MeV.  }\label{Fig:RealImagPotHTL}
\end{figure}

\begin{figure}[th!]
\centering
\includegraphics[scale=1.1]{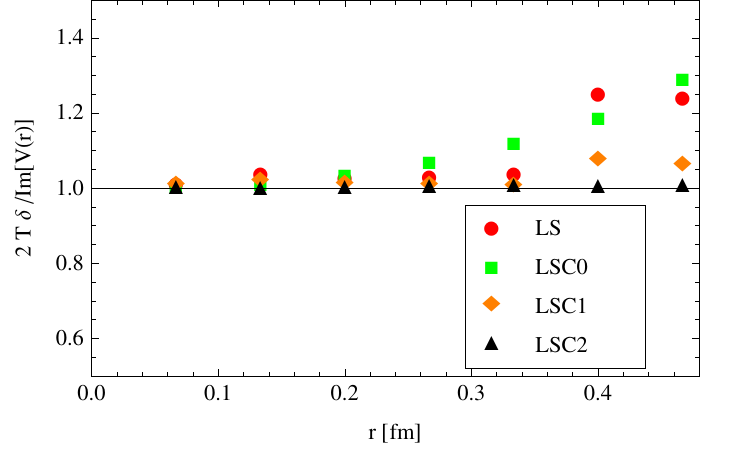}
\caption{ Visualization of the connection $\delta_\square=\frac{{\rm Im}[V]}{2T}$ between skewing parameter and imaginary part from the potential peak in the HTL Wilson loop spectrum. For the skewing to be correctly determined, background terms up to quadratic order need to be taken into account.}
\label{Fig:SqImagPotHTL}
\end{figure}

In the following we do not constrain any of the possible fitting parameters, i.e we allow e.g. $\Re[\sigma_\infty]$ and $\Im[V]$ to be determined separately by the fit. To estimate in which cases the use of improved fitting functions becomes necessary, we compare the results from a simple Lorentzian (L) (i.e. $\Re[\sigma_\infty]=0$ and $c_i=0$ in eq.\eqref{Eq:FitShapeFull}), a skewed Lorentzian (LS) (i.e. $c_i=0$ in eq.\eqref{Eq:FitShapeFull}) and the skewed Lorentzian with additional polynomial terms (LSC0, LSC1, LSC2)  (i.e. $c_{i>0}=0,~c_{i>1}=0,~c_{i>2}=0$ in eq.\eqref{Eq:FitShapeFull}).

We find that fitting with a simple Lorentzian (L) yields reasonable results only at very small separation distances $r<0.2\fm$, where the width of the peak itself does not exceed $100\MeV$. In this distance region, the use of a skewed Lorentzian (LS) improves the fit significantly and actually reproduces the peak shape quite well.

As shown in Fig.~\ref{Fig:SpecFits}, at separations of $r=0.49 \mathrm{\fm}$ the situation is already more involved, as the width of the peaks grows to around $150 \MeV$ and the shape deviates markedly from a naive Lorentzian. Adding the extra degree of freedom of skewing alone does not remedy the situation. Only after including the constant term (LSC0), arising from the early time variation of the function $\Phi(r,t)$, we find that that the spectral shape is reconstructed in an acceptable manner. Including the additional linear ($c_1$) and quadratic ($c_2$) coefficients improves the overall agreement with the spectral shape, while the extracted values of the peak position and width are unaffected. This stability against including higher order background terms gives us confidence in the reliability of the fit. 

After scrutinizing the goodness of fit, we can turn our attention to the actual values of the potential obtained in this manner. In Fig.~\ref{Fig:RealImagPotHTL} the values of the real- and imaginary part of the UV regularized potential $V^\Lambda(r)$ are shown in red (solid line) and the values obtained from the Wilson loop spectra fits are overlayed as discrete points. We find that the (LSC) fit successfully reproduces the real part of the potential at least up to the fourth digit. Note that the real part of the regularized potential shows an oscillating pattern absent in the UV complete $V(R)$, which is retraced by the (LSC) fit. In lattice QCD where both a UV and IR cutoff are present, similar oscillations might arise. 

As expected from the fitting of the spectral shapes, the determination of the real and imaginary part succeeds even for the naive Lorentzian as long as the width is below $100\MeV$. We find that the real part is less sensitive to the fitting function, only at larger distances, the Lorentzian overshoots the correct value, preventing us to observe the effect of Debye screening.

The values of the imaginary part show  a stronger dependence on the fitting function and the correct values are only obtained after including the constant term (LSC0). In particular at large distances $r>0.3\fm$ the simple Lorentzian and even the skewed Lorentzian overestimate the values of the imaginary part.

According to the relation in Eq.~\eqref{ImVSq} is should also be possible to extract the imaginary part of the potential through the skewing parameter $\delta_\square$. While for a correct determination of the skewing, a precise fit of the peak shape is necessary, it is indeed possible to use the (LSC) fit to successfully relate skewing and imaginary part of the potential as shown Fig.~\ref{Fig:SqImagPotHTL}.

We conclude hence that the extraction of both the real and imaginary part from Wilson loop spectra succeeds if the improved fitting function eq.\eqref{Eq:FitShapeFull} is deployed\footnote{We checked that by including the next higher order in eq.\eqref{Eq:FitShapeFull}, i.e. the term linear in frequency with $c_1\neq0$, improves the fit at larger $\omega>3.3{\rm \GeV}$ but does not change the extraction of the parameter values. If we go to higher temperatures, where the width becomes even larger or if we wish to fit the spectrum over a larger frequency interval around the peak, we will have to include higher terms of the $c_i$'s.} and therefore the knowledge of the shape of the lowest lying peak is sufficient to determine the potential. The results obtained with the (LSC) fit show a negligible deviation from the correct results and the deviation can be estimated by observing the variation of the fit results when 
introducing new 
fit parameters e.g. $c_1$. We also find that fitting the lowest peak with a simple Lorentzian leads to an overestimation of both the real and imaginary part of the potential which contributed to the counter-intuitive results of Ref.~\cite{Rothkopf:2011db}.

\section{MEM analysis of the perturbative Euclidean correlator}\label{sect6}

While the extraction of both real and imaginary part of the potential from the lowest lying peak structure in $\rho^\Lambda(r,\omega)$ has been shown to succeed in case of known HTL spectra in section \ref{sect5}, we now wish to face the numerically challenging aspect of actually reconstructing these spectra from a set of Euclidean time data points.

In the following we will deploy an MEM implementation with extended search space \cite{Rothkopf:2011ef} (for technical background see e.g. \cite{Jarrell1996133,Asakawa:2000tr,Jakovac:2006sf,Nickel:2006mm}) in an attempt to reconstruct from $N_\tau=32$ ideal imaginary time datapoints the most probable spectral function in the Bayesian sense. This number of available measurements along Euclidean time is representative for what we encounter in actual lattice QCD studies of correlation functions. By not adding additional noise and merely attaching artificial error bars to the correlators before feeding them to the MEM we deliberately choose the best case scenario in which any useful algorithm has to prevail\footnote{One reasoning behind our choice is that e.g. through the application of the multilevel algorithm it is possible to measure datapoints with very high accuracy by sacrificing a number of available datapoints.}.

\subsubsection{Wilson loop}

To choose appropriate parameters for the MEM we first inspect the Euclidean data points in the top graph of Fig.~\ref{2}. The strong suppression at small $\tau$ as well as the rise at $\tau\simeq\beta$ tell us that structures at large positive and negative frequencies contribute to the full spectrum. Thus we decide to discretize $\omega$ in an interval $I_\omega=[-126,189]\GeV$ by $N_\omega=800$ points using arithmetic with a precision of $384$ bits.

The necessity for a large negative value of $\omega_{\rm min}$, indicated by the data, implies that the $N_\tau$ basis functions in Bryan's search space do not contain enough variation to capture any peak at positive frequency. Hence we amend the search space by $48$ additional basis functions of the full search space, whose oscillations cover the whole range of $\omega$. The Levenberg-Marquardt (LM) algorithm is subsequently used to perform a search for the most probable spectral function within the confines of the above parameters.

\begin{figure}[t!]
\centering
\includegraphics[scale=0.17,angle=-90]{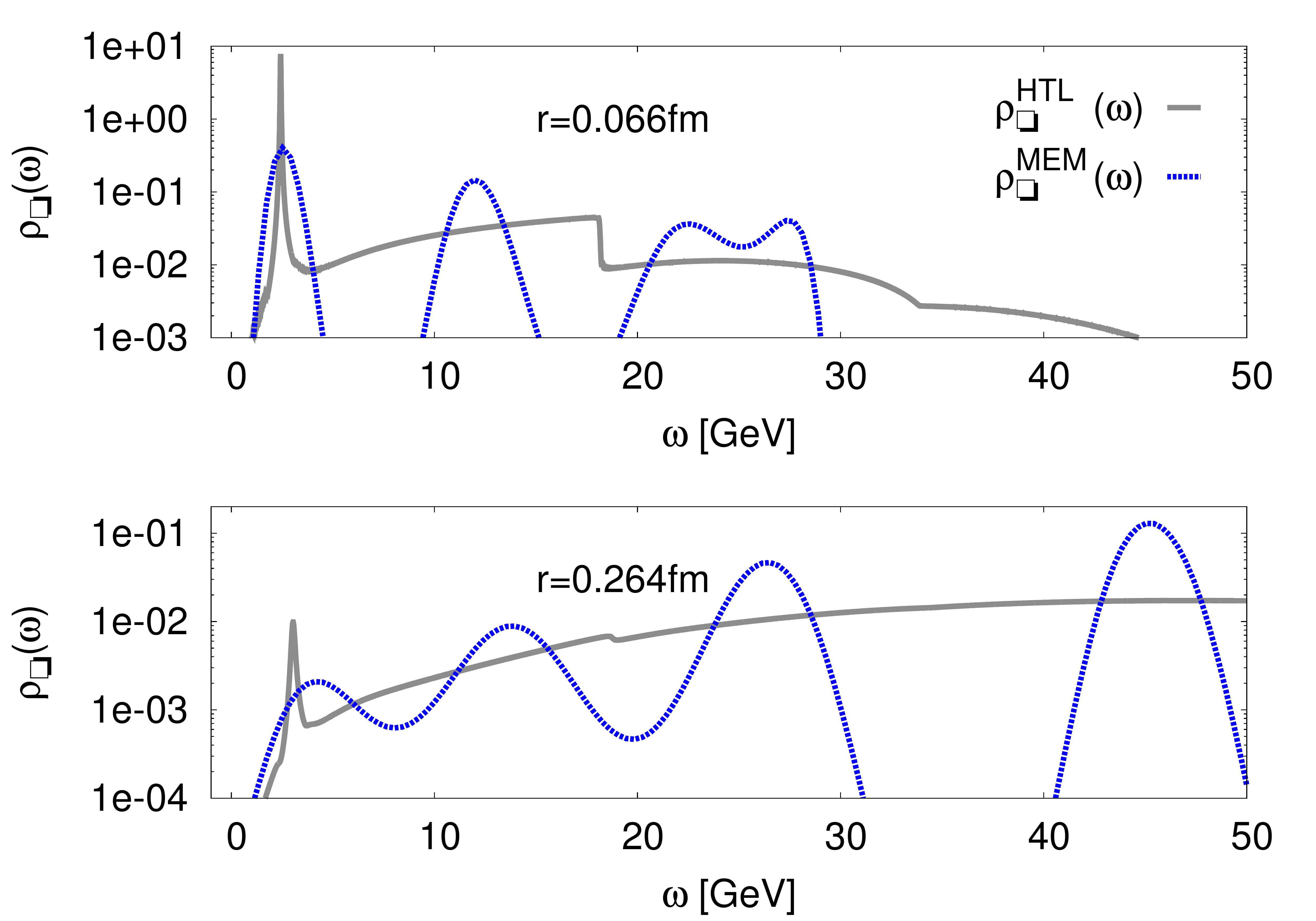}
 \caption{MEM reconstructed spectra (dashed, blue) at $r=0.066\fm$ (top) and $r=0.264\fm$ (bottom) based on $N_\tau=32$ ideal Euclidean HTL Wilson loop data points at $T=2.33T_C$. We discretize the frequency interval $I_\omega=[-126,189]\GeV$ by $N_\omega=800$ points and provide $N_{\rm ext}=N_\tau+48$ basis functions for the minimizer to reconstruct $\rho^{\rm MEM}_\square(\omega)$. The exact HTL result at the corresponding distance is given as gray solid curve. Note that even though the MEM recognizes the presence of the large background terms it fails to produce a smooth reconstruction. Both position and shape of the lowest lying peak are rather poorly captured, which we attribute in part to the presence of the large background contribution. The limited number of available degrees of freedom do not suffice to capture both small and large $(\omega>5\GeV)$ structures. }
 \label{Fig:MEMWLoopSpecCmp}
\end{figure}

Two of the resulting spectra are plotted in Fig.~\ref{Fig:MEMWLoopSpecCmp}. We find that while the presence of the large background is acknowledged by the MEM through several peaks  at frequencies $5\GeV<\omega<50\GeV$, it is at the same time difficult to obtain a good reconstruction of the lowest lying peak. At $r=0.066$ at least its position is captured satisfactorily, the width of the structure remains an order of magnitude too large.

\begin{figure}[t!]
\centering
\includegraphics[scale=1.2]{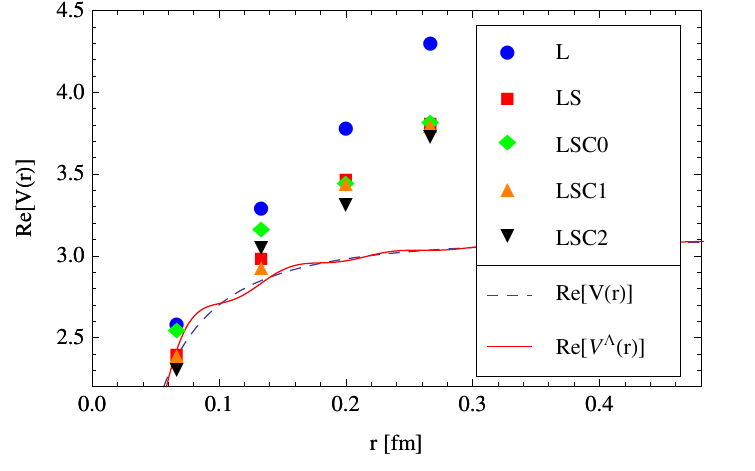}
 \caption{Real part of the potential extracted from the MEM reconstructed HTL Wilson loop spectrum at $T=2.33T_C$. We observe a consistent overestimation of the peak position, which persists even if higher background terms are included in the fitting function (e.g. LSC2). From the results of section \ref{sect5} it is apparent that this failure originates in a deficiency of the underlying MEM reconstructed spectra. }
 \label{Fig:WLoopMEMSpecReV}
\end{figure}

\begin{figure}[t!]
\centering
\includegraphics[scale=1.15]{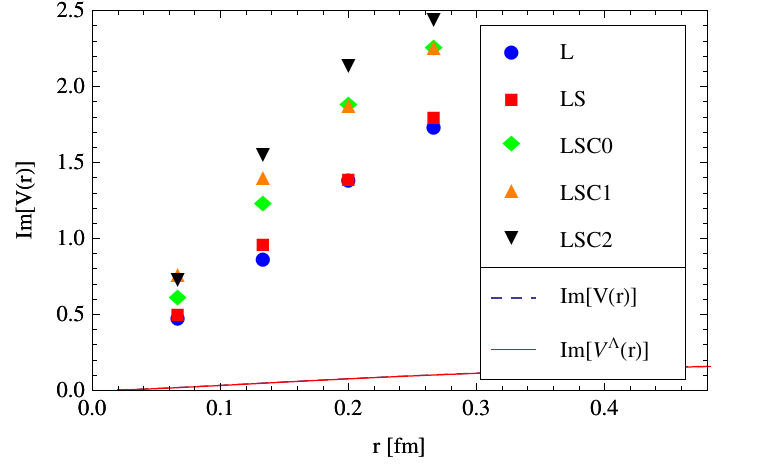}
 \caption{Imaginary part of the potential extracted from the MEM reconstructed HTL Wilson loop spectrum at $T=2.33T_C$. We observe a consistent overestimation of the peak width of more than one order of magnitude. Note that including more fitting parameters worsens the estimation of the peak, since the underlying spectra do not actually resemble a skewed Lorentzian.}
 \label{Fig:WLoopMEMSpecImV}
\end{figure}

Based on the MEM spectra we can proceed to fit the lowest peak using the fitting function Eq.~\eqref{Eq:FitShapeFull} analogous to the spectra of section \ref{sect5}. The results are given in Fig.~\ref{Fig:WLoopMEMSpecReV} and Fig.~\ref{Fig:WLoopMEMSpecImV}.  The inadequacy of the spectral reconstruction translates here into a consistent overestimation of the values for both real- and imaginary part of the potential. The shift in the peak position can be understood again from the presence of the large cusp divergence induced background, which together with the limited number of basis functions, pulls the peak towards higher $\omega$ in the reconstruction. Similar to observations in previous lattice QCD based studies, both real and imaginary part are of the same order of magnitude.

A few technical comments are in order. Our condition for finding the optimal spectrum in the Bayesian sense relies on a manual stopping criterion for the LM algorithm at relative improvements in the search of $\Delta=10^{-10}$ to limit the necessary time for one run to the order of days. This however is not yet a true minimum since the values meander around in tiny steps inside the search space without converging to a definite value within machine precision. This fact is sometimes reflected in non-smooth behavior of the $\alpha$ probability distribution.

Increasing the number of basis functions improves the reconstruction slightly, i.e. the width decreases, but even with $N_{\rm ext}>200$ we are not able to reproduce a Lorentzian peak shape including the characteristic tail structures. We furthermore only see marginal improvement in determining the lowest peak neither if the number of datapoints is increased nor if the size of the artificially attached errorbars is lowered. Since Bayesian inference is based on sound statistical reasoning with a well defined limit for infinitely many datapoints and ideal data these findings lead us to the conclusion that at this point it is not the properties of the supplied data but rather the implementation of the method that prevents us from a successful spectral reconstruction.

It should however be noted that the difficulty in correctly reconstructing the lowest peak is more difficult in the HTL case than what we expect to face on the lattice. Even though remnants of the momentum cutoff $\Lambda$ are found in the HTL spectrum, the fact that higher frequencies contribute to integrals within the HTL gluon spectral functions, used at intermediate steps of determining $W^{\rm HTL}_\square(\tau,r)$, allows the background to stretch far beyond our choice of $\Lambda=5\pi \GeV$. The presence of a sharp lattice cutoff would amputate such structures, the corresponding lattice correlator is less suppressed and thus the potential peak more easily reconstructed.

We arrive at a sobering conclusion. Based on the Maximum Entropy Method in its current form, even after including an extended search space, the reconstruction of the real and imaginary part from the Wilson loop is extremely challenging. One of the reasons is the presence of the large background structures introduced by the cusp divergences (see Fig.~\ref{Fig:MEMWLoopSpecCmp}), which furthermore suppress the amplitude of the lowest lying peak. All attempts at a reconstruction of a sharp peak at small $\omega$ are hampered since our limited reservoir of available degrees of freedom is depleted by structures not related to the physics of the potential. 

\subsubsection{Wilson line correlator}

The reason to investigate alternative observables such as the Wilson line correlator in Coulomb gauge as basis for an MEM reconstruction is now evident. As we have seen in section \ref{sect3} the absence of cusp divergences leads to a dramatically reduced suppression along the Euclidean time axis. The rise at $\tau=\beta$ observed in the Wilson loop is also virtually absent. This bodes well for an application of the standard MEM as the difficulties encountered in the previous subsection were directly connected with the divergences induced background contributions. 

\begin{figure}[t!]
\centering
\includegraphics[scale=0.17, angle=-90]{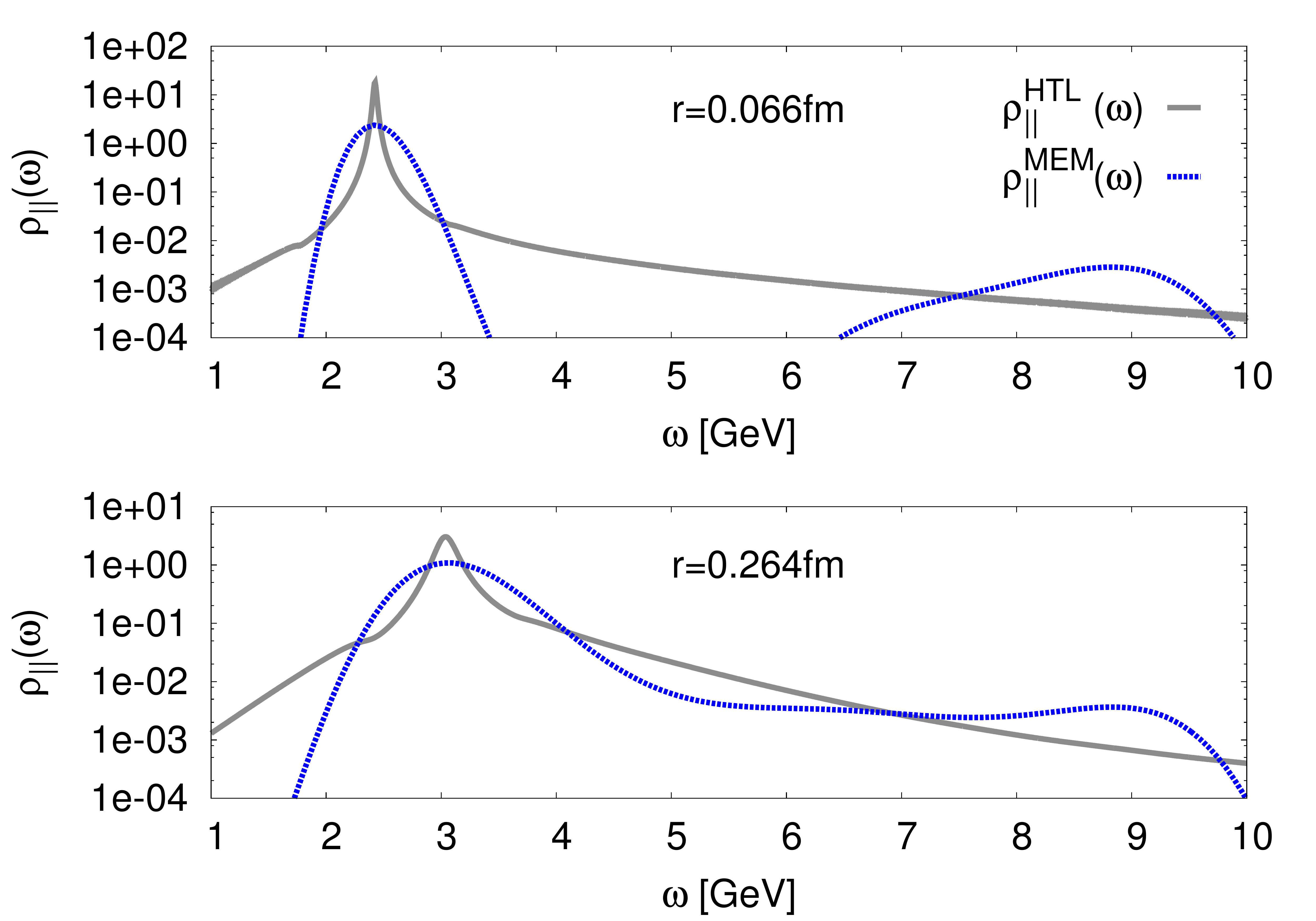}
 \caption{MEM reconstructed spectra (dashed, blue) at $r=0.066\fm$ (top) and $r=0.264\fm$ (bottom) based on $N_\tau=32$ ideal Euclidean HTL Wilson line correlator data points. We discretize the frequency interval $I_\omega=[-63,126]\GeV$ by $N_\omega=2500$ points and provide $N_{\rm ext}=N_\tau+48$ basis functions for the minimizer to reconstruct $\rho^{\rm MEM}_{||}(\omega)$. The exact HTL result at the corresponding distance is given as gray solid curve. Note the absence of large background terms which hampered the reconstruction in the Wilson loop case. While the peak position is captured in a satisfying manner, the width of the reconstructed peaks is almost two orders of magnitude too large}\label{Fig:MEMWLineSpecCmp}
\end{figure}

We choose to discretize frequencies in an interval $I_\omega=[-63,126]\GeV$ by $N_\omega=2500$ points. The different choice of frequency range and $N_\omega$ compared to the Wilson loop case reflects our expectation that the available degrees of freedom suffice to reconstruct a much more narrow lowest lying peak. The necessity for accommodating a large background is gone. 

\begin{figure}[t!]
\centering
\includegraphics[scale=1.1]{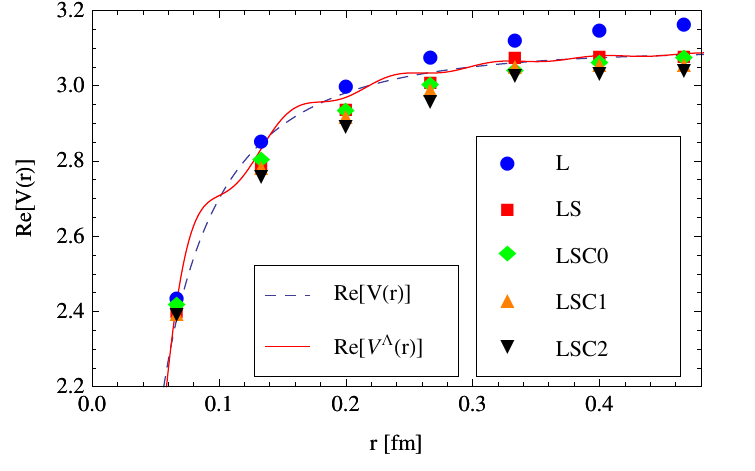}
 \caption{Real part of the potential extracted from the MEM reconstructed HTL Wilson line correlator spectrum. After inclusion of skewing a reasonable agreement with the HTL potential is obtained. Note however that the values at larger $r$ did not yet asymptote with respect to the inclusion of higher orders of background terms and tend to underestimate the correct values. The naive Lorentzian fit on the other hand leads to values that are too large.}\label{Fig:WLineMEMSpecReV}
\end{figure}

\begin{figure}[t!]
\centering
\includegraphics[scale=1.1]{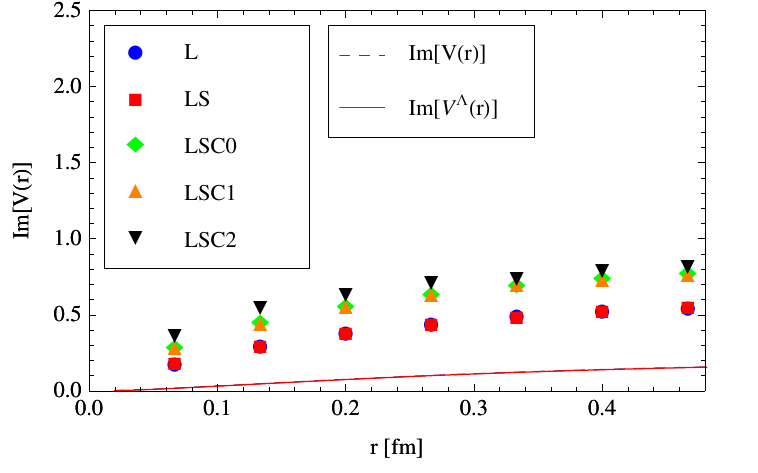}
 \caption{Imaginary part of the potential extracted from the MEM reconstructed HTL Wilson line correlator spectrum. Even though a much better resemblance of the reconstructed peaks with the exact HTL spectrum is obtained, we are still a factor five away from the actual values of the HTL imaginary part at this temperature.}\label{Fig:WLineMEMSpecImV}
\end{figure}

Fig.~\ref{Fig:MEMWLineSpecCmp} shows two of the resulting reconstructed spectra which exhibit a much better agreement with the correct HTL result than in the Wilson loop case. Note the factor five in the frequency axis compared Fig.~\ref{Fig:MEMWLoopSpecCmp}. Carrying out a fit with Eq.\ref{Eq:FitShapeFull} as in section \ref{sect5}, allows us to estimate the values of the real- and imaginary part of the potential shown in Fig.~\ref{Fig:WLineMEMSpecReV} and Fig.~\ref{Fig:WLineMEMSpecImV}. 

We observe that at least for the real part a reasonable agreement with the correct $\Re[V^{\rm HTL}]$ has been obtained, once the skewing is included (LS). What is striking however is that the the values for different fit functions (LSC0, LSC1, LSC2) do not yet seem to asymptote for larger separation distances and thus begin to underestimate the correct values. This behavior is connected to the fact that the shape of the reconstructed spectral peak does not resemble a skewed Lorentzian as can be seen in Fig.~\ref{Fig:MEMWLineSpecCmp}.

The estimation of the width of the peak still fares worse compared to the peak position even though the disagreement has been roughly reduced by a factor three. The absence of the divergences in the Wilson line correlator and hence the absence of background terms already leads to a much more narrow reconstructed peak compared to the Wilson loop scenario, it is however still not possible to reach the actual width of the exact result.

\begin{figure}[t!]
\centering
\includegraphics[scale=1.1]{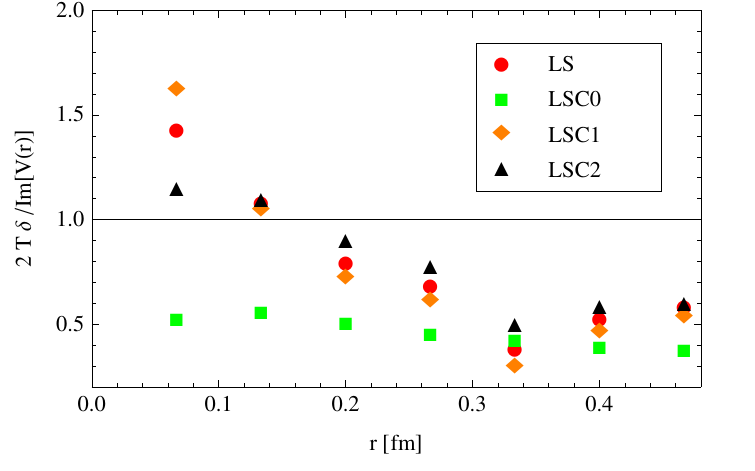}
 \caption{Test of the relation $\delta_{||}=\frac{{\rm Im}[V]}{2T}$ between the the fitted skewing parameter and the imaginary part. We find significant deviations from unity with no clear tendency of improvement. This tells us that at this stage, the reconstructed spectral shapes still do not reliably reproduce a skewed Lorentzian.}\label{Fig:WLineMEMSpecSkew}
\end{figure}

Since the reconstruction of the peak improved significantly for the Wilson line case, we wish to inspect, whether the relation between skewing and peak width is already visible in the MEM spectra. Fig.~\ref{Fig:WLineMEMSpecSkew} depicts the ratio between the fitting parameter $\delta^{\rm MEM}_{||}$ and ${\rm Im}[V]$ scaled by twice the temperature. The deviations from unity with no clear tendency of improvement tell us that at this stage, the reconstructed spectral shapes still do not reliably reproduce the skewed Lorentzian functional form actually encoded in the HTL Euclidean data. 

If the relation between skewing and imaginary part should turn out to hold beyond the leading order HTL approximation it would lend itself to checking the success of the MEM reconstruction. Note that in the Wilson loop case the extracted values of the skewing, besides being completely unstable between different fits did not show any correlation with the peak width.

Despite the obvious technical shortcomings, which hamper the numerical determination of the potential, our findings are encouraging in that they show how the choice of underlying observable can improve the chances for a successful extraction of the potential. 

At least in the leading order HTL approximation the late real-time physics content is the same whether the Wilson loop or the Wilson line correlator in Coulomb gauge is concerned. The absence of divergences in the latter however permits the MEM reconstruction to lie much closer to the correct values. 

From the point of view of lattice QCD practitioners, the effects of e.g. the smearing procedure on spatial links in the Wilson loop are connected with modifying physics near the UV cutoff and might therefore also lead to an improved observable with respect to the potential extraction.

\section{Conclusions and Outlook} \label{sect7}

The heavy quark correlator satisfies a Schr\"odinger equation at late real-times, parametrized by a complex potential. It has been proposed that this potential can in principle be extracted from the imaginary time Wilson loop measured on the lattice \cite{Rothkopf:2011db,Burnier:2012az}. The required steps involve an analytic continuation from Euclidean to Minkowski times, usually performed with the help of Bayesian inference and a fitting procedure that connects the position and shape of the lowest lying spectral peak to the real- and imaginary part of the potential respectively.

Using HTL perturbation theory, we performed a systematic check of the method with the aim to recover the well known potential of Ref.~\cite{Laine:2006ns}. At leading order all quantities relevant to the extraction of the potential can be calculated explicitly. After the introduction of a momentum cutoff for large spatial momenta, we were able to determine the full time dependence of both the Wilson loop and Coulomb gauge fixed Wilson line correlator in the Euclidean as well as Minkowskian setting. The discrete Fourier transform is used to calculate the corresponding spectral functions. 

We find that the expectations of Ref.~\cite{Burnier:2012az} are fulfilled, as all spectra contain a well defined lowest lying peak of skewed Lorentzian form. A major difference between Wilson loop and Wilson line correlator in Coulomb gauge is the presence of cusp divergences in the former, which translates into large background structures engulfing the potential peak. Nevertheless the exact same potential is encoded in the two observables at this order in HTL. 

A surprising fact is that the skewing parameter, related to the non-potential physics at early times, is the same for both observables and itself related to the imaginary part of the potential (see Eq.\eqref{VHTL}). It would be very interesting to study both whether both of these properties hold beyond the leading order HTL approximation or even on the lattice.

Based on the HTL spectra we checked whether fitting the functional form derived in \cite{Burnier:2012az} succeeds and found that indeed a very accurate determination of the potential is possible from exclusive knowledge of the lowest lying peak. We can hence replace late real-time limit in (\ref{Eq:DefPotSpec}) by fitting the low frequency realm of the spectrum. We find that the simple Lorentzian consistently overestimates the potential for intermediate and large values of the separation distance $r$, which offers a partial explanation for the large values observed in previous studies. By including skewing and higher order background terms from (\ref{Eq:FitShapeFull}) however we obtain very stable and reliable estimates for ${\rm Re}[V]$ and ${\rm Im}[V]$.

Our attempts of a standard MEM reconstruction of the spectra from ideal Euclidean time data-points revealed several challenges. We find that the large background, induced by the cusp divergences contained in the Wilson loop, makes a reconstruction very difficult. The broad structures not related to the potential absorb a large part of the limited number of degrees of freedom available to the MEM and prevent the potential peak to be captured in a satisfying manner. In addition we are not able to obtain a peak shape that resembles a skewed Lorentzian even after increasing the number of basis function for the MEM search space or adding additional datapoints. This leads us to conclude that it is not the properties of the supplied datapoints but instead the technical implementation of the MEM itself that prevents us from a successful reconstruction.

The situation is significantly improved in the case of the Wilson line correlator in Coulomb gauge, since the cusp divergences and hence the large background contributions in the spectrum are absent. Within the standard MEM it becomes possible to obtain a reasonable estimate of the real part of the potential, the imaginary part is still overestimated by at least a factor of five. It is still not possible to reconstruct the Lorentzian functional form encoded in the Euclidean correlator and thus the fits of (\ref{Eq:FitShapeFull}) to the both real and imaginary part become unstable if too many free parameters (e.g. $c_i$'s) are included. The favorable UV structure of the Wilson line correlator invites speculation whether it can also help us to determine the non-perturbative potential from lattice QCD.

At zero temperature, the potential is extracted from the Coulomb gauge Wilson lines anyway. In this case, it was shown in perturbation theory that at least to NNLO \cite{Schroder:1999sg} the full Wilson loop matches the Coulomb gauge Wilson lines. The general picture is that if a physical potential description exists at all in the large time limit, the details of the initial condition (how we close the Wilson loop) will not matter\footnote{unless, of course, one chooses a divergent gauge, for instance $A_0=0$}. 

At non-zero temperature, if the Euclidean correlator is considered, there is no meaning to a large time limit as $\tau<\beta$. However if we consider the analytically continued real time correlator or its spectrum, the infinite time limit might be considered similarly to the way it is performed at zero temperature. We showed here that the leading order HTL results agree one might expect, as in the zero temperature case, that the large time limit of the correlators do not to depend on the way we close them at the boundary. Should this assumption hold, then the potential can be extracted from the Wilson line spectrum (or the smeared Wilson loop) as well, although not directly from the Euclidean data, for which the infinite time limit does not make sense.

\acknowledgments

The authors thank N.~Brambilla, M.~Escobedo, J.~Ghiglieri,  M.~Laine, P.~Petreczky and A.~Vario for very helpful discussions and the DFG-Heisenberg group of Y. Schr\"oder at Bielefeld University for generous access to computational resources. This work was partly supported by the Swiss National Science Foundation (SNF) under grant 200021-140234. Y.~B. acknowledges partial support by the European Community under the FP7 programme HadronPhysics3.

\appendix

\section{HTL spectral functions}\label{A}

Inverting the relation (\ref{spectralf}), we have,
\beq
\rho(q^0)=\frac{\Delta(q_0\to -i(q^0+i\e))-\Delta(q_0\to -i(q^0-i\e))}{2i},
\eeq
with $\Delta(q_0)= q_0^2+\bq^2+\Pi(q_0,\bq)$. The spectral functions have a pole (gluon for the transverse, plasmon for the longitudinal) in the region $q^0>q$ and a continuous part in the region $q^0<q$. They are antisymmetric in $q^0$ and we restrict the formulas below to $q^0>0$.
Explicitly, the electric spectral function reads
\bea
\rho_L(q^0,q)&\overset{q^0>q}{=}& \pi\delta\left[f_L(q_0)\right],\\
\rho_L(q^0,q)&\overset{q>q^0>0}{=}&\frac{-\frac{\pi \mD^2}{2}\notag \frac{q^0 q}{q^2-(q^0)^2}}{\left[q^2+\mD^2\left(1-l(q^0,q)\right)\right]^2+\left[\frac{\pi\mD^2}{2}\frac{q^0}{q}\right]^2},\label{rhoL}
\eea
with
\bea
l(q^0,q)&=&\frac{q^0}{2q}\ln\left(\frac{q^0+q}{q-q^0}\right),\notag\\
f_L(q_0)&=&\left((q^0)^2-q^2\right)\left[1+\frac{\mD^2}{q^2}\left(1-l(q,q^0)\right)\right]\notag
\eea
For the transverse spectral function, we have
\bea
&&\rho_T(q^0,q)\overset{q^0>q}{=} \pi\delta\left[f_T(q^0)\right],\\\notag
&&\rho_T(q^0,q)\overset{q>q^0>0}{=}\\&&\quad\frac{\pi\frac{\mD^2q^0q}{4(q^2-(q^0)^2)}}{\left[\left(q^2+\frac{\mD^2}{2}l(q^0,q)\right)+\frac{\mD^2 (q^0)^2}{2(q^2-(q^0)^2)}\right]^2+ \left[\frac{\pi\mD^2q^0}{4q}\right]^2}\notag
\eea
with
\beq f_T(q^0)=\left((q^0)^2-q^2\right)\left[1+\frac{\mD^2}{2 q^2}l(q,q^0)\right]-\frac{(q^0)^2\mD^2}{2q^3}\notag
\eeq

\section{Convergence of the pole contributions}\label{pole}
For the longitudinal spectral function, the solution to the delta function behave at large $q^0$ as 
$$
q^0_L\simeq\left[1+2\exp\left(-2\frac{q^2+\mD^2}{\mD^2}\right)\right].
$$
The full integral is then convergent because the factor  
$$\left(1-\frac{q^2}{(q^0_L)^2}\right)$$
is exponentially small.

For the transverse spectral function, at large $q$, the pole is sitting at
\beq
q^0_T\sim\sqrt{q^2+\frac{\mD^2}{2}+\mathcal{O}\left(\frac{\mD^2}{g^2} \log\frac{\mD^2}{g^2}\right)}.\label{q0T}
\eeq
The contribution from the poles is basically given by integrating with respect to $q^0$ using the $\delta$-function 
$$\int dq^0\delta(f_{L,T}(q^0))\to |f_{L,T}'(q^0)|^{-1}\sim 1/q,$$ 
which gives in the transverse case a linear divergence.

\section{Cusp divergence}\label{C}

We calculate here the divergence coming form the pole of the transverse HTL spectral function (\ref{31}) in dimensional regularization and show that it matches the cusp divergences.

For $\tau=0,~\beta$ the function $h(\tau,q^0_T)$ vanishes and hence we have no contribution, note that there are no cusps in these degenerate cases. For $\tau\neq 0,~\beta$, one can decompose the factor
$$n_B(q_T^0)h(\tau,q^0_T)=1-\frac{e^{\tau q^0_T}-e^{(\beta-\tau) q^0_T}}{e^{\beta q^0_T-1}}$$
where the second term lead to a convergent integral. 

Using equation (\ref{q0T}), the remaining part of the integrand in (\ref{31}) can also be decomposed as
\bea
\frac{2-\frac{\sin(q r)}{qr}-cos(q r)- q r Si(qr)}{|f'_T(q_T^0)|}\notag\\=\frac{1}{\sqrt{q^2+\mD^2/2}}-\frac \pi 4 r+ \mathrm{finite}\notag
\eea
where the terms omitted are UV and IR finite.
Using dimensional regularization, the linear divergent term drops and only the $\tau$ and $r$ independent term survives,
\beq
W^{(2)}_{pole,T}(\tau,r)=\frac{C_F g^2}{2\pi^2\epsilon}+\mathrm{finite}.\label{cuspdiv}
\eeq
The cusp multiplicative divergence arising from one angle $\gamma$ reads \cite{Korchemsky:1987wg, Brandt:1982gz}
$$Z_{cusp}= 1+\frac{g^2 C_F}{8\pi^2 \epsilon}(1+(\pi-\gamma)\cot \gamma)$$
In the Wilson loop case \cite{Berwein:2012mw} with four angles $\gamma=\pi/2$, we get as leading order contribution precisely the divergent part of (\ref{cuspdiv}).

\bibliography{references2.bib}

\begin{thebibliography}{10}%
\makeatletter
\providecommand \@ifxundefined [1]{%
 \ifx #1\undefined \expandafter \@firstoftwo
 \else \expandafter \@secondoftwo
\fi
}%
\providecommand \@ifnum [1]{%
 \ifnum #1\expandafter \@firstoftwo
 \else \expandafter \@secondoftwo
\fi
}%
\providecommand \enquote [1]{``#1''}%
\providecommand \bibnamefont  [1]{#1}%
\providecommand \bibfnamefont [1]{#1}%
\providecommand \citenamefont [1]{#1}%
\providecommand\href[0]{\@sanitize\@href}%
\providecommand\@href[1]{\endgroup\@@startlink{#1}\endgroup\@@href}%
\providecommand\@@href[1]{#1\@@endlink}%
\providecommand \@sanitize [0]{\begingroup\catcode`\&12\catcode`\#12\relax}%
\@ifxundefined \pdfoutput {\@firstoftwo}{%
 \@ifnum{\z@=\pdfoutput}{\@firstoftwo}{\@secondoftwo}%
}{%
 \providecommand\@@startlink[1]{\leavevmode\special{html:<a href="#1">}}%
 \providecommand\@@endlink[0]{\special{html:</a>}}%
}{%
 \providecommand\@@startlink[1]{%
  \leavevmode
  \pdfstartlink
   attr{/Border[0 0 1 ]/H/I/C[0 1 1]}%
   user{/Subtype/Link/A<</Type/Action/S/URI/URI(#1)>>}%
  \relax
 }%
 \providecommand\@@endlink[0]{\pdfendlink}%
}%
\providecommand \url  [0]{\begingroup\@sanitize \@url }%
\providecommand \@url [1]{\endgroup\@href {#1}{\urlprefix}}%
\providecommand \urlprefix [0]{URL }%
\providecommand \Eprint[0]{\href }%
\@ifxundefined \urlstyle {%
  \providecommand \doi [1]{doi:\discretionary{}{}{}#1}%
}{%
  \providecommand \doi [0]{doi:\discretionary{}{}{}\begingroup
  \urlstyle{rm}\Url }%
}%
\providecommand \doibase [0]{http://dx.doi.org/}%
\providecommand \Doi[1]{\href{\doibase#1}}%
\providecommand \bibAnnote [3]{%
  \BibitemShut{#1}%
  \begin{quotation}\noindent
    \textsc{Key:}\ #2\\\textsc{Annotation:}\ #3%
  \end{quotation}%
}%
\providecommand \bibAnnoteFile [2]{%
  \IfFileExists{#2}{\bibAnnote {#1} {#2} {\input{#2}}}{}%
}%
\providecommand \typeout [0]{\immediate \write \m@ne }%
\providecommand \selectlanguage [0]{\@gobble}%
\providecommand \bibinfo [0]{\@secondoftwo}%
\providecommand \bibfield [0]{\@secondoftwo}%
\providecommand \translation [1]{[#1]}%
\providecommand \BibitemOpen[0]{}%
\providecommand \bibitemStop [0]{}%
\providecommand \bibitemNoStop [0]{.\EOS\space}%
\providecommand \EOS [0]{\spacefactor3000\relax}%
\providecommand \BibitemShut [1]{\csname bibitem#1\endcsname}%
\bibitem{Matsui:1986dk}%
  \BibitemOpen
  \bibfield{author}{%
  \bibinfo {author} {\bibfnamefont{T.}~\bibnamefont{Matsui}}\ and\ \bibinfo
  {author} {\bibfnamefont{H.}~\bibnamefont{Satz}},\ }%
  \bibfield{journal}{%
  \Doi{10.1016/0370-2693(86)91404-8}{\bibinfo {journal} {Phys.Lett.}}\ }%
  \textbf{\bibinfo {volume} {B178}},\ \bibinfo {pages} {416} (\bibinfo {year}
  {1986})%
  \bibAnnoteFile{NoStop}{Matsui:1986dk}%
\bibitem{Adare:2006ns}%
  \BibitemOpen
  \bibfield{author}{%
  \bibinfo {author} {\bibfnamefont{A.}~\bibnamefont{Adare}} \emph{et~al.}
  (\bibinfo {collaboration} {PHENIX Collaboration}),\ }%
  \bibfield{journal}{%
  \Doi{10.1103/PhysRevLett.98.232301}{\bibinfo {journal} {Phys.Rev.Lett.}}\ }%
  \textbf{\bibinfo {volume} {98}},\ \bibinfo {pages} {232301} (\bibinfo {year}
  {2007}),\
  \Eprint{http://arxiv.org/abs/nucl-ex/0611020}{arXiv:nucl-ex/0611020}%
  \bibAnnoteFile{NoStop}{Adare:2006ns}%
\bibitem{Tang:2011kr}%
  \BibitemOpen
  \bibfield{author}{%
  \bibinfo {author} {\bibfnamefont{Z.}~\bibnamefont{Tang}} (\bibinfo
  {collaboration} {STAR Collaboration}),\ }%
  \bibfield{journal}{%
  \Doi{10.1088/0954-3899/38/12/124107}{\bibinfo {journal} {J.Phys.G}}\ }%
  \textbf{\bibinfo {volume} {G38}},\ \bibinfo {pages} {124107} (\bibinfo {year}
  {2011}),\ \Eprint{http://arxiv.org/abs/1107.0532}{arXiv:1107.0532}%
  \bibAnnoteFile{NoStop}{Tang:2011kr}%
\bibitem{Chatrchyan:2012np}%
  \BibitemOpen
  \bibfield{author}{%
  \bibinfo {author} {\bibfnamefont{S.}~\bibnamefont{Chatrchyan}} \emph{et~al.}
  (\bibinfo {collaboration} {CMS Collaboration}),\ }%
  \bibfield{journal}{%
  \Doi{10.1007/JHEP05(2012)063}{\bibinfo {journal} {JHEP}}\ }%
  \textbf{\bibinfo {volume} {1205}},\ \bibinfo {pages} {063} (\bibinfo {year}
  {2012}),\ \Eprint{http://arxiv.org/abs/1201.5069}{arXiv:1201.5069}%
  \bibAnnoteFile{NoStop}{Chatrchyan:2012np}%
\bibitem{Abelev:2012rv}%
  \BibitemOpen
  \bibfield{author}{%
  \bibinfo {author} {\bibfnamefont{B.}~\bibnamefont{Abelev}} \emph{et~al.}
  (\bibinfo {collaboration} {ALICE Collaboration})}%
   (\bibinfo {year} {2012}),\
  \Eprint{http://arxiv.org/abs/1202.1383}{arXiv:1202.1383}%
  \bibAnnoteFile{NoStop}{Abelev:2012rv}%
\bibitem{Brambilla:2004jw}%
  \BibitemOpen
  \bibfield{author}{%
  \bibinfo {author} {\bibfnamefont{N.}~\bibnamefont{Brambilla}}, \bibinfo
  {author} {\bibfnamefont{A.}~\bibnamefont{Pineda}}, \bibinfo {author}
  {\bibfnamefont{J.}~\bibnamefont{Soto}},\ and\ \bibinfo {author}
  {\bibfnamefont{A.}~\bibnamefont{Vairo}},\ }%
  \bibfield{journal}{%
  \Doi{10.1103/RevModPhys.77.1423}{\bibinfo {journal} {Rev.Mod.Phys.}}\ }%
  \textbf{\bibinfo {volume} {77}},\ \bibinfo {pages} {1423} (\bibinfo {year}
  {2005}),\ \Eprint{http://arxiv.org/abs/hep-ph/0410047}{arXiv:hep-ph/0410047}%
  \bibAnnoteFile{NoStop}{Brambilla:2004jw}%
\bibitem{Bazavov:2012ka}%
  \BibitemOpen
  \bibfield{author}{%
  \bibinfo {author} {\bibfnamefont{A.}~\bibnamefont{Bazavov}}, \bibinfo
  {author} {\bibfnamefont{N.}~\bibnamefont{Brambilla}}, \bibinfo {author}
  {\bibfnamefont{X.}~\bibnamefont{Garcia~i Tormo}}, \bibinfo {author}
  {\bibfnamefont{P.}~\bibnamefont{Petreczky}}, \bibinfo {author}
  {\bibfnamefont{J.}~\bibnamefont{Soto}}, \emph{et~al.},\ }%
  \bibfield{journal}{%
  \Doi{10.1103/PhysRevD.86.114031}{\bibinfo {journal} {Phys.Rev.}}\ }%
  \textbf{\bibinfo {volume} {D86}},\ \bibinfo {pages} {114031} (\bibinfo {year}
  {2012}),\ \Eprint{http://arxiv.org/abs/1205.6155}{arXiv:1205.6155}%
  \bibAnnoteFile{NoStop}{Bazavov:2012ka}%
\bibitem{Laine:2006ns}%
  \BibitemOpen
  \bibfield{author}{%
  \bibinfo {author} {\bibfnamefont{M.}~\bibnamefont{Laine}}, \bibinfo {author}
  {\bibfnamefont{O.}~\bibnamefont{Philipsen}}, \bibinfo {author}
  {\bibfnamefont{P.}~\bibnamefont{Romatschke}},\ and\ \bibinfo {author}
  {\bibfnamefont{M.}~\bibnamefont{Tassler}},\ }%
  \bibfield{journal}{%
  \Doi{10.1088/1126-6708/2007/03/054}{\bibinfo {journal} {JHEP}}\ }%
  \textbf{\bibinfo {volume} {0703}},\ \bibinfo {pages} {054} (\bibinfo {year}
  {2007}),\ \Eprint{http://arxiv.org/abs/hep-ph/0611300}{arXiv:hep-ph/0611300}%
  \bibAnnoteFile{NoStop}{Laine:2006ns}%
\bibitem{Brambilla:2008cx}%
  \BibitemOpen
  \bibfield{author}{%
  \bibinfo {author} {\bibfnamefont{N.}~\bibnamefont{Brambilla}}, \bibinfo
  {author} {\bibfnamefont{J.}~\bibnamefont{Ghiglieri}}, \bibinfo {author}
  {\bibfnamefont{A.}~\bibnamefont{Vairo}},\ and\ \bibinfo {author}
  {\bibfnamefont{P.}~\bibnamefont{Petreczky}},\ }%
  \bibfield{journal}{%
  \Doi{10.1103/PhysRevD.78.014017}{\bibinfo {journal} {Phys.Rev.}}\ }%
  \textbf{\bibinfo {volume} {D78}},\ \bibinfo {pages} {014017} (\bibinfo {year}
  {2008}),\ \Eprint{http://arxiv.org/abs/0804.0993}{arXiv:0804.0993}%
  \bibAnnoteFile{NoStop}{Brambilla:2008cx}%
\bibitem{Beraudo:2007ky}%
  \BibitemOpen
  \bibfield{author}{%
  \bibinfo {author} {\bibfnamefont{A.}~\bibnamefont{Beraudo}}, \bibinfo
  {author} {\bibfnamefont{J.-P.}\ \bibnamefont{Blaizot}},\ and\ \bibinfo
  {author} {\bibfnamefont{C.}~\bibnamefont{Ratti}},\ }%
  \bibfield{journal}{%
  \Doi{10.1016/j.nuclphysa.2008.03.001}{\bibinfo {journal} {Nucl.Phys.}}\ }%
  \textbf{\bibinfo {volume} {A806}},\ \bibinfo {pages} {312} (\bibinfo {year}
  {2008}),\ \Eprint{http://arxiv.org/abs/0712.4394}{arXiv:0712.4394}%
  \bibAnnoteFile{NoStop}{Beraudo:2007ky}%
\bibitem{Brambilla:2010vq}%
  \BibitemOpen
  \bibfield{author}{%
  \bibinfo {author} {\bibfnamefont{N.}~\bibnamefont{Brambilla}}, \bibinfo
  {author} {\bibfnamefont{M.~A.}\ \bibnamefont{Escobedo}}, \bibinfo {author}
  {\bibfnamefont{J.}~\bibnamefont{Ghiglieri}}, \bibinfo {author}
  {\bibfnamefont{J.}~\bibnamefont{Soto}},\ and\ \bibinfo {author}
  {\bibfnamefont{A.}~\bibnamefont{Vairo}},\ }%
  \bibfield{journal}{%
  \Doi{10.1007/JHEP09(2010)038}{\bibinfo {journal} {JHEP}}\ }%
  \textbf{\bibinfo {volume} {1009}},\ \bibinfo {pages} {038} (\bibinfo {year}
  {2010}),\ \Eprint{http://arxiv.org/abs/1007.4156}{arXiv:1007.4156}%
  \bibAnnoteFile{NoStop}{Brambilla:2010vq}%
\bibitem{Brambilla:2011mk}%
  \BibitemOpen
  \bibfield{author}{%
  \bibinfo {author} {\bibfnamefont{N.}~\bibnamefont{Brambilla}}, \bibinfo
  {author} {\bibfnamefont{M.~A.}\ \bibnamefont{Escobedo}}, \bibinfo {author}
  {\bibfnamefont{J.}~\bibnamefont{Ghiglieri}},\ and\ \bibinfo {author}
  {\bibfnamefont{A.}~\bibnamefont{Vairo}},\ }%
  \bibfield{journal}{%
  \Doi{10.1007/JHEP07(2011)096}{\bibinfo {journal} {JHEP}}\ }%
  \textbf{\bibinfo {volume} {1107}},\ \bibinfo {pages} {096} (\bibinfo {year}
  {2011}),\ \Eprint{http://arxiv.org/abs/1105.4807}{arXiv:1105.4807 [hep-ph]}%
  \bibAnnoteFile{NoStop}{Brambilla:2011mk}%
\bibitem{Rothkopf:2011db}%
  \BibitemOpen
  \bibfield{author}{%
  \bibinfo {author} {\bibfnamefont{A.}~\bibnamefont{Rothkopf}}, \bibinfo
  {author} {\bibfnamefont{T.}~\bibnamefont{Hatsuda}},\ and\ \bibinfo {author}
  {\bibfnamefont{S.}~\bibnamefont{Sasaki}},\ }%
  \bibfield{journal}{%
  \Doi{10.1103/PhysRevLett.108.162001}{\bibinfo {journal} {Phys.Rev.Lett.}}\ }%
  \textbf{\bibinfo {volume} {108}},\ \bibinfo {pages} {162001} (\bibinfo {year}
  {2012}),\ \Eprint{http://arxiv.org/abs/1108.1579}{arXiv:1108.1579}%
  \bibAnnoteFile{NoStop}{Rothkopf:2011db}%
\bibitem{Digal:2005ht}%
  \BibitemOpen
  \bibfield{author}{%
  \bibinfo {author} {\bibfnamefont{S.}~\bibnamefont{Digal}}, \bibinfo {author}
  {\bibfnamefont{O.}~\bibnamefont{Kaczmarek}}, \bibinfo {author}
  {\bibfnamefont{F.}~\bibnamefont{Karsch}},\ and\ \bibinfo {author}
  {\bibfnamefont{H.}~\bibnamefont{Satz}},\ }%
  \bibfield{journal}{%
  \Doi{10.1140/epjc/s2005-02309-7}{\bibinfo {journal} {Eur.Phys.J.}}\ }%
  \textbf{\bibinfo {volume} {C43}},\ \bibinfo {pages} {71} (\bibinfo {year}
  {2005}),\ \Eprint{http://arxiv.org/abs/hep-ph/0505193}{arXiv:hep-ph/0505193}%
  \bibAnnoteFile{NoStop}{Digal:2005ht}%
\bibitem{Burnier:2012az}%
  \BibitemOpen
  \bibfield{author}{%
  \bibinfo {author} {\bibfnamefont{Y.}~\bibnamefont{Burnier}}\ and\ \bibinfo
  {author} {\bibfnamefont{A.}~\bibnamefont{Rothkopf}},\ }%
  \bibfield{journal}{%
  \Doi{10.1103/PhysRevD.86.051503}{\bibinfo {journal} {Phys.Rev.}}\ }%
  \textbf{\bibinfo {volume} {D86}},\ \bibinfo {pages} {051503} (\bibinfo {year}
  {2012}),\ \Eprint{http://arxiv.org/abs/1208.1899}{arXiv:1208.1899}%
  \bibAnnoteFile{NoStop}{Burnier:2012az}%
\bibitem{Brambilla:2013dpa}%
  \BibitemOpen
  \bibfield{author}{%
  \bibinfo {author} {\bibfnamefont{N.}~\bibnamefont{Brambilla}}, \bibinfo
  {author} {\bibfnamefont{M.~A.}\ \bibnamefont{Escobedo}}, \bibinfo {author}
  {\bibfnamefont{J.}~\bibnamefont{Ghiglieri}},\ and\ \bibinfo {author}
  {\bibfnamefont{A.}~\bibnamefont{Vairo}}}%
   (\bibinfo {year} {2013}),\
  \Eprint{http://arxiv.org/abs/1303.6097}{arXiv:1303.6097}%
  \bibAnnoteFile{NoStop}{Brambilla:2013dpa}%
\bibitem{Rothkopf:2009pk}%
  \BibitemOpen
  \bibfield{author}{%
  \bibinfo {author} {\bibfnamefont{A.}~\bibnamefont{Rothkopf}}, \bibinfo
  {author} {\bibfnamefont{T.}~\bibnamefont{Hatsuda}},\ and\ \bibinfo {author}
  {\bibfnamefont{S.}~\bibnamefont{Sasaki}},\ }%
  \bibfield{journal}{%
  \bibinfo {journal} {PoS}\ }%
  \textbf{\bibinfo {volume} {LAT2009}},\ \bibinfo {pages} {162} (\bibinfo
  {year} {2009}),\ \Eprint{http://arxiv.org/abs/0910.2321}{arXiv:0910.2321}%
  \bibAnnoteFile{NoStop}{Rothkopf:2009pk}%
\bibitem{Burnier:2007qm}%
  \BibitemOpen
  \bibfield{author}{%
  \bibinfo {author} {\bibfnamefont{Y.}~\bibnamefont{Burnier}}, \bibinfo
  {author} {\bibfnamefont{M.}~\bibnamefont{Laine}},\ and\ \bibinfo {author}
  {\bibfnamefont{M.}~\bibnamefont{Vepsalainen}},\ }%
  \bibfield{journal}{%
  \Doi{10.1088/1126-6708/2008/01/043}{\bibinfo {journal} {JHEP}}\ }%
  \textbf{\bibinfo {volume} {0801}},\ \bibinfo {pages} {043} (\bibinfo {year}
  {2008}),\ \Eprint{http://arxiv.org/abs/0711.1743}{arXiv:0711.1743}%
  \bibAnnoteFile{NoStop}{Burnier:2007qm}%
\bibitem{Burnier:2008ia}%
  \BibitemOpen
  \bibfield{author}{%
  \bibinfo {author} {\bibfnamefont{Y.}~\bibnamefont{Burnier}}, \bibinfo
  {author} {\bibfnamefont{M.}~\bibnamefont{Laine}},\ and\ \bibinfo {author}
  {\bibfnamefont{M.}~\bibnamefont{Vepsalainen}},\ }%
  \bibfield{journal}{%
  \Doi{10.1088/1126-6708/2009/02/008}{\bibinfo {journal} {JHEP}}\ }%
  \textbf{\bibinfo {volume} {0902}},\ \bibinfo {pages} {008} (\bibinfo {year}
  {2009}),\ \Eprint{http://arxiv.org/abs/0812.2105}{arXiv:0812.2105}%
  \bibAnnoteFile{NoStop}{Burnier:2008ia}%
\bibitem{Ding:2012iy}%
  \BibitemOpen
  \bibfield{author}{%
  \bibinfo {author} {\bibfnamefont{H.-T.}\ \bibnamefont{Ding}}, \bibinfo
  {author} {\bibfnamefont{A.}~\bibnamefont{Francis}}, \bibinfo {author}
  {\bibfnamefont{O.}~\bibnamefont{Kaczmarek}}, \bibinfo {author}
  {\bibfnamefont{F.}~\bibnamefont{Karsch}}, \bibinfo {author}
  {\bibfnamefont{H.}~\bibnamefont{Satz}}, \emph{et~al.}}%
   (\bibinfo {year} {2012}),\
  \Eprint{http://arxiv.org/abs/1210.0292}{arXiv:1210.0292}%
  \bibAnnoteFile{NoStop}{Ding:2012iy}%
\bibitem{Ding:2012sp}%
  \BibitemOpen
  \bibfield{author}{%
  \bibinfo {author} {\bibfnamefont{H.}~\bibnamefont{Ding}}, \bibinfo {author}
  {\bibfnamefont{A.}~\bibnamefont{Francis}}, \bibinfo {author}
  {\bibfnamefont{O.}~\bibnamefont{Kaczmarek}}, \bibinfo {author}
  {\bibfnamefont{F.}~\bibnamefont{Karsch}}, \bibinfo {author}
  {\bibfnamefont{H.}~\bibnamefont{Satz}}, \emph{et~al.},\ }%
  \bibfield{journal}{%
  \Doi{10.1103/PhysRevD.86.014509}{\bibinfo {journal} {Phys.Rev.}}\ }%
  \textbf{\bibinfo {volume} {D86}},\ \bibinfo {pages} {014509} (\bibinfo {year}
  {2012}),\ \Eprint{http://arxiv.org/abs/1204.4945}{arXiv:1204.4945}%
  \bibAnnoteFile{NoStop}{Ding:2012sp}%
\bibitem{Cuniberti:2001hm}%
  \BibitemOpen
  \bibfield{author}{%
  \bibinfo {author} {\bibfnamefont{G.}~\bibnamefont{Cuniberti}}, \bibinfo
  {author} {\bibfnamefont{E.}~\bibnamefont{De~Micheli}},\ and\ \bibinfo
  {author} {\bibfnamefont{G.~A.}\ \bibnamefont{Viano}},\ }%
  \bibfield{journal}{%
  \Doi{10.1007/s002200000324}{\bibinfo {journal} {Commun.Math.Phys.}}\ }%
  \textbf{\bibinfo {volume} {216}},\ \bibinfo {pages} {59} (\bibinfo {year}
  {2001}),\
  \Eprint{http://arxiv.org/abs/cond-mat/0109175}{arXiv:cond-mat/0109175}%
  \bibAnnoteFile{NoStop}{Cuniberti:2001hm}%
\bibitem{Burnier:2011jq}%
  \BibitemOpen
  \bibfield{author}{%
  \bibinfo {author} {\bibfnamefont{Y.}~\bibnamefont{Burnier}}, \bibinfo
  {author} {\bibfnamefont{M.}~\bibnamefont{Laine}},\ and\ \bibinfo {author}
  {\bibfnamefont{L.}~\bibnamefont{Mether}},\ }%
  \bibfield{journal}{%
  \Doi{10.1140/epjc/s10052-011-1619-0}{\bibinfo {journal} {Eur.Phys.J.}}\ }%
  \textbf{\bibinfo {volume} {C71}},\ \bibinfo {pages} {1619} (\bibinfo {year}
  {2011}),\ \Eprint{http://arxiv.org/abs/1101.5534}{arXiv:1101.5534}%
  \bibAnnoteFile{NoStop}{Burnier:2011jq}%
\bibitem{Burnier:2012ts}%
  \BibitemOpen
  \bibfield{author}{%
  \bibinfo {author} {\bibfnamefont{Y.}~\bibnamefont{Burnier}}\ and\ \bibinfo
  {author} {\bibfnamefont{M.}~\bibnamefont{Laine}},\ }%
  \bibfield{journal}{%
  \Doi{10.1140/epjc/s10052-012-1902-8}{\bibinfo {journal} {Eur.Phys.J.}}\ }%
  \textbf{\bibinfo {volume} {C72}},\ \bibinfo {pages} {1902} (\bibinfo {year}
  {2012}),\ \Eprint{http://arxiv.org/abs/1201.1994}{arXiv:1201.1994}%
  \bibAnnoteFile{NoStop}{Burnier:2012ts}%
\bibitem{Viano1}%
  \BibitemOpen
  \bibfield{author}{%
  \bibinfo {author} {\bibfnamefont{E.}~\bibnamefont{De~Micheli}}\ and\ \bibinfo
  {author} {\bibfnamefont{G.~A.}\ \bibnamefont{Viano}},\ }%
  \bibfield{journal}{%
  \bibinfo {journal} {J. Math. Anal. Appl.}\ }%
  \textbf{\bibinfo {volume} {246}},\ \bibinfo {pages} {520} (\bibinfo {year}
  {2000}),\ \Eprint{http://arxiv.org/abs/math/0512029}{arXiv:math/0512029}%
  \bibAnnoteFile{NoStop}{Viano1}%
\bibitem{Viano2}%
  \BibitemOpen
  \bibfield{author}{%
  \bibinfo {author} {\bibfnamefont{E.}~\bibnamefont{De~Micheli}}\ and\ \bibinfo
  {author} {\bibfnamefont{G.~A.}\ \bibnamefont{Viano}},\ }%
  \bibfield{journal}{%
  \Doi{10.1007/s002200000324}{\bibinfo {journal} {J. Math. Anal. Appl.}}\ }%
  \textbf{\bibinfo {volume} {234}},\ \bibinfo {pages} {265} (\bibinfo {year}
  {1999}),\ \Eprint{http://arxiv.org/abs/math/0511604}{arXiv:math/0511604}%
  \bibAnnoteFile{NoStop}{Viano2}%
\bibitem{Burnier:2009bk}%
  \BibitemOpen
  \bibfield{author}{%
  \bibinfo {author} {\bibfnamefont{Y.}~\bibnamefont{Burnier}}, \bibinfo
  {author} {\bibfnamefont{M.}~\bibnamefont{Laine}},\ and\ \bibinfo {author}
  {\bibfnamefont{M.}~\bibnamefont{Vepsalainen}},\ }%
  \bibfield{journal}{%
  \Doi{10.1007/JHEP01(2010)054}{\bibinfo {journal} {JHEP}}\ }%
  \textbf{\bibinfo {volume} {1001}},\ \bibinfo {pages} {054} (\bibinfo {year}
  {2010}),\ \Eprint{http://arxiv.org/abs/0911.3480}{arXiv:0911.3480}%
  \bibAnnoteFile{NoStop}{Burnier:2009bk}%
\bibitem{Berwein:2012mw}%
  \BibitemOpen
  \bibfield{author}{%
  \bibinfo {author} {\bibfnamefont{M.}~\bibnamefont{Berwein}}, \bibinfo
  {author} {\bibfnamefont{N.}~\bibnamefont{Brambilla}}, \bibinfo {author}
  {\bibfnamefont{J.}~\bibnamefont{Ghiglieri}},\ and\ \bibinfo {author}
  {\bibfnamefont{A.}~\bibnamefont{Vairo}},\ }%
  \bibfield{journal}{%
  \Doi{10.1007/JHEP03(2013)069}{\bibinfo {journal} {JHEP}}\ }%
  \textbf{\bibinfo {volume} {1303}},\ \bibinfo {pages} {069} (\bibinfo {year}
  {2013}),\ \Eprint{http://arxiv.org/abs/1212.4413}{arXiv:1212.4413}%
  \bibAnnoteFile{NoStop}{Berwein:2012mw}%
\bibitem{Korchemsky:1987wg}%
  \BibitemOpen
  \bibfield{author}{%
  \bibinfo {author} {\bibfnamefont{G.}~\bibnamefont{Korchemsky}}\ and\ \bibinfo
  {author} {\bibfnamefont{A.}~\bibnamefont{Radyushkin}},\ }%
  \bibfield{journal}{%
  \Doi{10.1016/0550-3213(87)90277-X}{\bibinfo {journal} {Nucl.Phys.}}\ }%
  \textbf{\bibinfo {volume} {B283}},\ \bibinfo {pages} {342} (\bibinfo {year}
  {1987})%
  \bibAnnoteFile{NoStop}{Korchemsky:1987wg}%
\bibitem{Brandt:1982gz}%
  \BibitemOpen
  \bibfield{author}{%
  \bibinfo {author} {\bibfnamefont{R.~A.}\ \bibnamefont{Brandt}}, \bibinfo
  {author} {\bibfnamefont{A.}~\bibnamefont{Gocksch}}, \bibinfo {author}
  {\bibfnamefont{M.}~\bibnamefont{Sato}},\ and\ \bibinfo {author}
  {\bibfnamefont{F.}~\bibnamefont{Neri}},\ }%
  \bibfield{journal}{%
  \Doi{10.1103/PhysRevD.26.3611}{\bibinfo {journal} {Phys.Rev.}}\ }%
  \textbf{\bibinfo {volume} {D26}},\ \bibinfo {pages} {3611} (\bibinfo {year}
  {1982})%
  \bibAnnoteFile{NoStop}{Brandt:1982gz}%
\bibitem{Bazavov:2013zha}%
  \BibitemOpen
  \bibfield{author}{%
  \bibinfo {author} {\bibfnamefont{A.}~\bibnamefont{Bazavov}}\ and\ \bibinfo
  {author} {\bibfnamefont{P.}~\bibnamefont{Petreczky}}}%
   (\bibinfo {year} {2013}),\
  \Eprint{http://arxiv.org/abs/1303.5500}{arXiv:1303.5500}%
  \bibAnnoteFile{NoStop}{Bazavov:2013zha}%
\bibitem{Maezawa:2011aa}%
  \BibitemOpen
  \bibfield{author}{%
  \bibinfo {author} {\bibfnamefont{Y.}~\bibnamefont{Maezawa}}, \bibinfo
  {author} {\bibfnamefont{T.}~\bibnamefont{Umeda}}, \bibinfo {author}
  {\bibfnamefont{S.}~\bibnamefont{Aoki}}, \bibinfo {author}
  {\bibfnamefont{S.}~\bibnamefont{Ejiri}}, \bibinfo {author}
  {\bibfnamefont{T.}~\bibnamefont{Hatsuda}}, \emph{et~al.},\ }%
  \bibfield{journal}{%
  \Doi{10.1143/PTP.128.955}{\bibinfo {journal} {Prog.Theor.Phys.}}\ }%
  \textbf{\bibinfo {volume} {128}},\ \bibinfo {pages} {955} (\bibinfo {year}
  {2012}),\ \Eprint{http://arxiv.org/abs/1112.2756}{arXiv:1112.2756}%
  \bibAnnoteFile{NoStop}{Maezawa:2011aa}%
\bibitem{Kaczmarek:2005ui}%
  \BibitemOpen
  \bibfield{author}{%
  \bibinfo {author} {\bibfnamefont{O.}~\bibnamefont{Kaczmarek}}\ and\ \bibinfo
  {author} {\bibfnamefont{F.}~\bibnamefont{Zantow}},\ }%
  \bibfield{journal}{%
  \Doi{10.1103/PhysRevD.71.114510}{\bibinfo {journal} {Phys.Rev.}}\ }%
  \textbf{\bibinfo {volume} {D71}},\ \bibinfo {pages} {114510} (\bibinfo {year}
  {2005}),\
  \Eprint{http://arxiv.org/abs/hep-lat/0503017}{arXiv:hep-lat/0503017}%
  \bibAnnoteFile{NoStop}{Kaczmarek:2005ui}%
\bibitem{Luscher:1996ug}%
  \BibitemOpen
  \bibfield{author}{%
  \bibinfo {author} {\bibfnamefont{M.}~\bibnamefont{Luscher}}, \bibinfo
  {author} {\bibfnamefont{S.}~\bibnamefont{Sint}}, \bibinfo {author}
  {\bibfnamefont{R.}~\bibnamefont{Sommer}}, \bibinfo {author}
  {\bibfnamefont{P.}~\bibnamefont{Weisz}},\ and\ \bibinfo {author}
  {\bibfnamefont{U.}~\bibnamefont{Wolff}},\ }%
  \bibfield{journal}{%
  \Doi{10.1016/S0550-3213(97)00080-1}{\bibinfo {journal} {Nucl.Phys.}}\ }%
  \textbf{\bibinfo {volume} {B491}},\ \bibinfo {pages} {323} (\bibinfo {year}
  {1997}),\
  \Eprint{http://arxiv.org/abs/hep-lat/9609035}{arXiv:hep-lat/9609035}%
  \bibAnnoteFile{NoStop}{Luscher:1996ug}%
\bibitem{Aoyama:1981ev}%
  \BibitemOpen
  \bibfield{author}{%
  \bibinfo {author} {\bibfnamefont{S.}~\bibnamefont{Aoyama}},\ }%
  \bibfield{journal}{%
  \Doi{10.1016/0550-3213(82)90023-2}{\bibinfo {journal} {Nucl.Phys.}}\ }%
  \textbf{\bibinfo {volume} {B194}},\ \bibinfo {pages} {513} (\bibinfo {year}
  {1982})%
  \bibAnnoteFile{NoStop}{Aoyama:1981ev}%
\bibitem{Rothkopf:2011ef}%
  \BibitemOpen
  \bibfield{author}{%
  \bibinfo {author} {\bibfnamefont{A.}~\bibnamefont{Rothkopf}},\ }%
  \bibfield{journal}{%
  \Doi{10.1016/j.jcp.2012.12.023}{\bibinfo {journal} {J.Comput.Phys.}}\ }%
  \textbf{\bibinfo {volume} {238}},\ \bibinfo {pages} {106} (\bibinfo {year}
  {2013}),\ \Eprint{http://arxiv.org/abs/1110.6285}{arXiv:1110.6285}%
  \bibAnnoteFile{NoStop}{Rothkopf:2011ef}%
\bibitem{Jarrell1996133}%
  \BibitemOpen
  \bibfield{author}{%
  \bibinfo {author} {\bibfnamefont{M.}~\bibnamefont{Jarrell}}\ and\ \bibinfo
  {author} {\bibfnamefont{J.}~\bibnamefont{Gubernatis}},\ }%
  \bibfield{journal}{%
  \Doi{10.1016/0370-1573(95)00074-7}{\bibinfo {journal} {Physics Reports}}\ }%
  \textbf{\bibinfo {volume} {269}},\ \bibinfo {pages} {133} (\bibinfo {year}
  {1996}),\ ISSN \bibinfo {issn} {0370-1573}%
  \bibAnnoteFile{NoStop}{Jarrell1996133}%
\bibitem{Asakawa:2000tr}%
  \BibitemOpen
  \bibfield{author}{%
  \bibinfo {author} {\bibfnamefont{M.}~\bibnamefont{Asakawa}}, \bibinfo
  {author} {\bibfnamefont{T.}~\bibnamefont{Hatsuda}},\ and\ \bibinfo {author}
  {\bibfnamefont{Y.}~\bibnamefont{Nakahara}},\ }%
  \bibfield{journal}{%
  \Doi{10.1016/S0146-6410(01)00150-8}{\bibinfo {journal}
  {Prog.Part.Nucl.Phys.}}\ }%
  \textbf{\bibinfo {volume} {46}},\ \bibinfo {pages} {459} (\bibinfo {year}
  {2001})%
  \bibAnnoteFile{NoStop}{Asakawa:2000tr}%
\bibitem{Jakovac:2006sf}%
  \BibitemOpen
  \bibfield{author}{%
  \bibinfo {author} {\bibfnamefont{A.}~\bibnamefont{Jakovac}}, \bibinfo
  {author} {\bibfnamefont{P.}~\bibnamefont{Petreczky}}, \bibinfo {author}
  {\bibfnamefont{K.}~\bibnamefont{Petrov}},\ and\ \bibinfo {author}
  {\bibfnamefont{A.}~\bibnamefont{Velytsky}},\ }%
  \bibfield{journal}{%
  \Doi{10.1103/PhysRevD.75.014506}{\bibinfo {journal} {Phys.Rev.}}\ }%
  \textbf{\bibinfo {volume} {D75}},\ \bibinfo {pages} {014506} (\bibinfo {year}
  {2007}),\
  \Eprint{http://arxiv.org/abs/hep-lat/0611017}{arXiv:hep-lat/0611017}%
  \bibAnnoteFile{NoStop}{Jakovac:2006sf}%
\bibitem{Nickel:2006mm}%
  \BibitemOpen
  \bibfield{author}{%
  \bibinfo {author} {\bibfnamefont{D.}~\bibnamefont{Nickel}},\ }%
  \bibfield{journal}{%
  \Doi{10.1016/j.aop.2006.09.002}{\bibinfo {journal} {Annals Phys.}}\ }%
  \textbf{\bibinfo {volume} {322}},\ \bibinfo {pages} {1949} (\bibinfo {year}
  {2007}),\ \Eprint{http://arxiv.org/abs/hep-ph/0607224}{arXiv:hep-ph/0607224}%
  \bibAnnoteFile{NoStop}{Nickel:2006mm}%
\bibitem{Schroder:1999sg}%
  \BibitemOpen
  \bibfield{author}{%
  \bibinfo {author} {\bibfnamefont{Y.}~\bibnamefont{Schroder}},\ }%
  \bibfield{journal}{%
  \bibinfo {journal} {PhD thesis~}}%
   (\bibinfo {year} {1999})%
  \bibAnnoteFile{NoStop}{Schroder:1999sg}%
\end{thebibliography}%
  
\end{document}